\documentclass[acmtog]{acmart} 
\AtBeginDocument{%
  }

\setcopyright{acmlicensed}
\copyrightyear{2025}
\acmYear{2025}
\acmDOI{10.1145/3757377.3763947}
\acmConference[SA Conference Papers ’25]{SIGGRAPH Asia' 25 Conference Papers}{December 15--18,
  2025}{Hong Kong, Hong Kong}
\acmISBN{979-8-4007-2137-3/2025/12}

\acmSubmissionID{139}

\usepackage{multirow}
\usepackage{caption}
\usepackage{hyperref}
\usepackage{dashrule}
\usepackage{placeins}

\begin{document}

\title{PartComposer: Learning and Composing Part-Level Concepts from Single-Image Examples}

\author{Junyu Liu}
\email{ljunyu381@gmail.com}
\email{liu_junyu@alumini.brown.edu}
\email{junyu.liu@epfl.ch}
\orcid{0009-0006-9309-5871}
\affiliation{%
  \institution{Brown University}
  \city{Providence}
  \state{Rhode Island}
  \country{USA}
}
\affiliation{%
  \institution{École Polytechnique Fédérale de Lausanne}
  \city{Lausanne}
  \state{Vaud}
  \country{Switzerland}
}

\author{R. Kenny Jones}
\email{russell\_jones@brown.edu}
\affiliation{%
  \institution{Brown University}
  \city{Providence}
  \state{Rhode Island}
  \country{USA}
}
\affiliation{%
  \institution{Stanford University}
  \city{Stanford}
  \state{California}
  \country{USA}
}

\author{Daniel Ritchie}
\email{daniel\_ritchie@brown.edu}
\affiliation{%
  \institution{Brown University}
  \city{Providence}
  \state{Rhode Island}
  \country{USA}
}

\renewcommand{\shortauthors}{Liu et al.}

\begin{abstract}

We present PartComposer: a framework for part-level concept learning from single-image examples that enables text-to-image diffusion models to compose novel objects from meaningful components. Existing methods either struggle with effectively learning fine-grained concepts or require a large dataset as input. We propose a dynamic data synthesis pipeline generating diverse part compositions to address one-shot data scarcity. Most importantly, we propose to maximize the mutual information between denoised latents and structured concept codes via a concept predictor, enabling direct regulation on concept disentanglement and re-composition supervision. Our method achieves strong disentanglement and controllable composition, outperforming subject and part-level baselines when mixing concepts from the same, or different, object categories.  Our code is released in \url{https://github.com/Junyu-Liu-Nate/partcomposer}.

\end{abstract}


\ccsdesc[500]{Computing methodologies~Machine learning; Computer graphics}

\keywords{personalization, multiple concept extraction, information theory}

\begin{teaserfigure}
  \includegraphics[width=\textwidth]{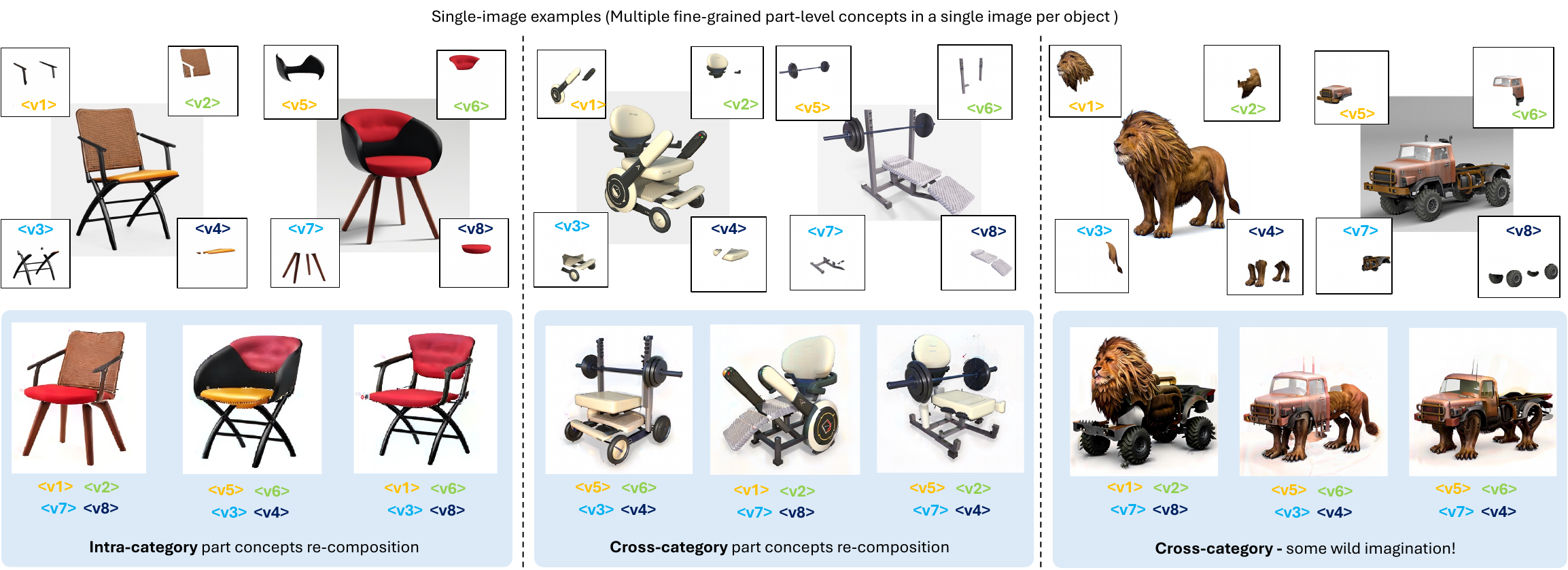}
  \caption{Given single-image examples where each image contains multiple fine-grained parts, our pipeline can learn descriptive concepts for these parts with clear disentanglement and flexibly re-compose them to generate new objects, for both intra-category and cross-category objects.}
  \Description{}
  \label{fig:teaser}
\end{teaserfigure}

\maketitle

\section{Introduction}
\label{sec:intro}

Visually inspired and creative generation emerges from the ability to compose new objects from familiar parts~\cite{24_Conceptlab, 25_PiT}. From virtual creatures to fantastical designs, part-level concept composition is a powerful paradigm for visual imagination. For instance, given images of several example chairs (Fig.~\ref{fig:teaser}), designers could imagine composing a new chair that contains parts borrowed across the example chairs and is structurally realistic. However, there are two major challenges for this task. Firstly, the \textbf{part-level} concepts are more fine-grained than subject-level concepts, and require structural information to be composed into reasonable objects. This makes it more challenging to properly learn the identity of concepts and achieve clear disentanglement between different concepts. Secondly, without dedicated 3D models or multi-view images of objects, most real-world examples in concept learning and composition are \textbf{single-image examples} (i.e., there is just one image per object). Such scarce data results in very poor data variety (i.e., very limited ground truth part composition), challenging the capability of generative models to properly compose unseen combinations.  

\begin{figure}[t]
    \centering
    \includegraphics[width=\linewidth]{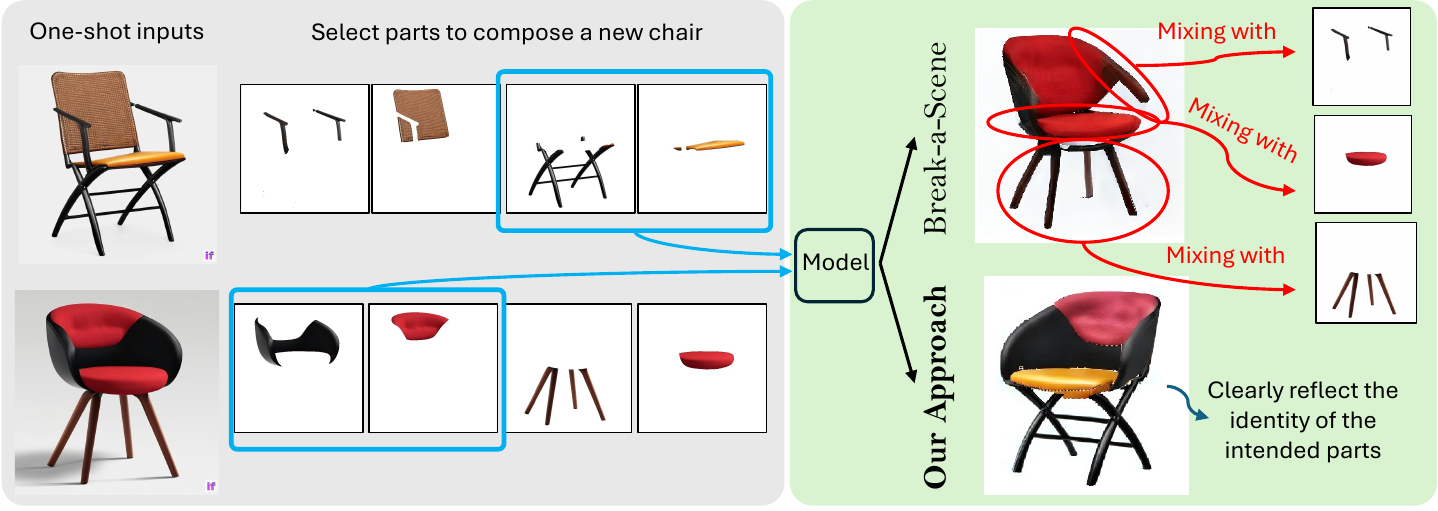}
    \caption{Illustration of the challenge in learning fine-grained concepts from single-image examples. Break-a-Scene~\cite{23_Break-a-Scene} results in multiple entanglements while our method cleanly reflects target concepts.}
    \label{fig:concept_mixing_illustration}
\end{figure}

A commonly used approach in concept learning is to extract visual concepts from generative models in the form of latent codes~\cite{22_TI,23_DreamBooth,23_CustomDiffusion,23_SVDiff,23_ELITE,23_Taming,24_InstantBooth}. Each of the extracted concepts encodes the identity of an image object (e.g., a red seat cushion) and can be used with other concepts in creative image generation~\cite{23_Break-a-Scene,23_PartCraft,24_Conceptlab,24_MuDI}. In this context, text-to-image diffusion models serve as a versatile tool for learning compositional concepts through personalization. A common paradigm is to learn new or specialized token embeddings for concepts, through fine-tuning the diffusion model (updating the embeddings and/or the model weights) via latent diffusion loss. The learned concept token embeddings can be directly used in prompts at inference time to generate images containing the target concepts. Several recent methods have explored dealing with single-image inputs~\cite{23_ELITE, 23_Taming, 24_InstantBooth, 23_Break-a-Scene, 25_TokenVerse}, but they operate at the \textit{subject level}. Other methods focus on part-level concept learning~\cite{23_PartCraft, 25_PiT}, but they require a dataset of images as input. A common problem is that these methods tend to fail to disentangle and retain the identity of fine-grained parts. Although several works~\cite{23_Break-a-Scene, 23_PartCraft} propose to use cross-attention loss between concept tokens and the visual contents to disentangle concepts, our experiments show that this method alone cannot effectively deal with part-level concepts from one-shot inputs. Fig.~\ref{fig:concept_mixing_illustration} illustrates this problem when composing 4 parts across 2 chairs, where some part-level concepts are ambiguous or even missing in composing new part combinations.

To overcome these challenges, we make two critical observations and propose corresponding modifications. Firstly, proper augmentation on the possible part combinations should be explored, and the structural integrity of objects should be preserved in the generative model prior. We propose a \textbf{dynamic data synthesis pipeline} to generate rich part-level supervision from single-image examples, by grouping union-sampled input images with randomly synthesized part combination images. Secondly, we observe that there is a lack of regulation of the information encoded in different concepts, resulting in entangled and ambiguous concept learning and composition. We propose a \textbf{mutual information maximization framework} that explicitly aligns denoised latents with structured part-level concept codes, by inferring which concepts are preserved given denoised latents. To effectively formalize such regulation, we argue that penalizing both wrong compositions and wrong localization of concepts are essential to achieve good concept disentanglement. Thus, we introduce a concept predictor that performs both classification and segmentation on latent features, regularizing the embedding space to reflect concept presence and spatial structure. 

We evaluate our methods on single-image part composition tasks from the same object categories and across different categories. We compare the part composition results with several baselines~\cite{23_Break-a-Scene, 23_PartCraft, 24_MuDI, 25_PiT} in terms of concept preserving capability and visual quality. Our method significantly out-performs baselines in preserving the correct identity of intended part-level concepts, while maintaining image quality. When given cross-category examples, our methods can produce interesting part composition of virtual objects with reasonable structure (Fig.~\ref{fig:teaser}), enabling great artistic imagination. We also provide an ablation study for our key designs and demonstrate the flexibility of our methods on different input forms. 

In summary, our contributions are: (1) We propose to focus on a challenging task, learning part-level concepts from single-image inputs, which relaxes the input requirement for objects' images. (2) We propose a dynamic data synthesis pipeline to automatically augment the scarce one-shot inputs and balance between structural integrity and part combination variety. (3) We propose a maximizing information scheme to explicitly disentangle the information encoded in concepts and supervise composition through a concept predictor with classification and segmentation.
\section{Background in Visual Concept Learning}
\label{append:sec:background}

\paragraph{Input Requirements}
Pioneering works like Textural Inversion~\cite{22_TI} and DreamBooth~\cite{23_DreamBooth} in visual concept learning require multiple images as input to encode a single concept. Several follow-up works have loosened the input constraint as learning a single concept from single-image examples~\cite{23_ELITE, 23_Taming, 24_InstantBooth}. Multi-concept learning methods from multiple inputs are also being discovered~\cite{23_CustomDiffusion, 23_SVDiff}. Break-a-Scene~\cite{23_Break-a-Scene} first proposes a pipeline to learn multiple subject-level concepts from single images, enabling great flexibility in re-compose the learned concepts.

\paragraph{Concept Granularity}
Most concept learning methods focus on learning concepts for the entire image or on the subject level~\cite{23_ELITE, 23_Taming, 24_InstantBooth, 23_CustomDiffusion, 23_SVDiff, 23_Break-a-Scene, 24_Image_Multi_Words, 24_ConceptExpress, 24_MuDI, 25_TokenVerse}. The target use case for these methods is to generate images that are organically composed by the required subject(s) according to user-specified prompts. However, to enable more creativity in generating new subjects, part-level concept learning is required. This more fine-grained concept learning task has been explored by PartCraft~\cite{23_PartCraft}. They learn a large dictionary of concepts for different parts of creatures and generate new virtual creatures that have not been seen in the real world. However, their method relies on training on a large dataset and struggles with single-image examples. Piece-it-Together (PiT)~\cite{25_PiT} also targets part-level concept learning,  by directly operating in a carefully chosen IP-Adapter+~\cite{23_Ip-adapter} representation space and synthesizes a complete and coherent concept by training a generative model to fill in missing information conditioned on a strong domain-specific prior. Unlike optimization-heavy approaches, PiT enables efficient inference, supports diverse sampling from sparse inputs, and allows flexible semantic manipulation. However, PiT requires training on class-specific datasets, where single-image examples are not supported and generalization to out-of-distribution data is limited.

\paragraph{Common Challenges}
A common challenge in multi-concept learning is to disentangle the identity of concepts~\cite{23_Break-a-Scene, 23_PartCraft, 24_MuDI}. To achieve good disentanglement, dynamic masking of different compositions of concepts and using cross-attention mechanisms to regulate the diffusion model fine-tuning are commonly used in methods like Break-a-Scene~\cite{23_Break-a-Scene} and PartCraft~\cite{23_PartCraft}. MuDI~\cite{24_MuDI} proposes a dynamic concept composition method and mean-shifted inference technique to improve the decoupling of different concepts. However, concepts missing or inaccurate identity still occur in all of these methods when re-composing concepts in the generative process~\cite{23_Break-a-Scene, 23_PartCraft, 24_MuDI}, especially when dealing with 4 or more concepts. Our observation is that all existing works do not provide any explicit regulation on the information encoded in different concepts and the composition process, causing inaccurate multi-concept remixing. Other common challenges include learning a large number of concepts (e.g., more than 4) from single inputs~\cite{23_Break-a-Scene}.
\section{Method}
\label{sec:method}

Given single-image examples, we aim to learn part concepts and arbitrarily re-compose them from each input to generate new objects. Our pipeline build upon the commonly used diffusion model customization approaches (introduced in Section~\ref{sec:intro}) like Break-a-Scene~\cite{23_Break-a-Scene} and PartCraft~\cite{23_PartCraft}, where standard diffusion loss $\mathcal{L_{\text{ldm}}}$ and cross-attention losses $\mathcal{L_{\text{attn}}}$ are used to learn and disentangle concept tokens. However, to tackle the challenges of learning part-level concepts from single-image examples, we need to deal with the extreme data scarcity and provide proper supervision on part-level information. To this end, we propose a dynamic data synthesis method to augment the limited data, and a maximizing mutual information scheme to enable clear disentanglement of parts' concepts and good composition capabilities for the generative model. Fig.~\ref{fig:method-pipeline} gives an overview of our approach.

We would like to emphasize that, although our pipeline is based on text-to-image diffusion models, we treat part-level concepts as purely visual information and do not require semantics or annotations for these part concepts. The tokens and text prompts only serve as media of templates to encode and convey such information.

\begin{figure}[t]
    \centering
    \includegraphics[width=\linewidth]{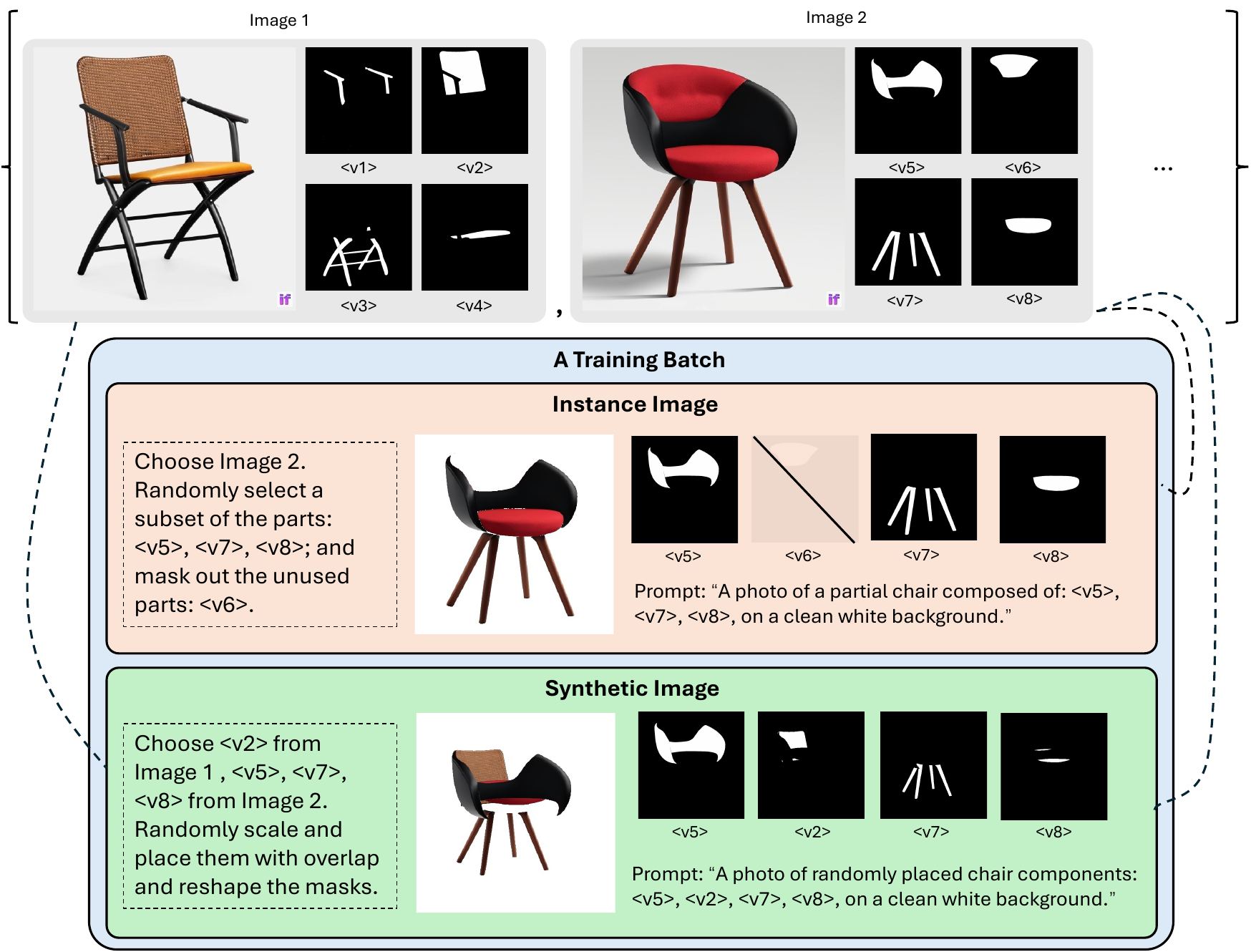}
    \caption{Dynamic data synthesis. Each training batch includes an instance image with masked parts and a synthetic image with randomly sampled and placed parts from multiple inputs, enabling diverse part-level supervision from single-image examples.}
    \label{fig:method-data-synth}
\end{figure}

\begin{figure*}[h]
    \centering
    \includegraphics[width=0.95\linewidth]{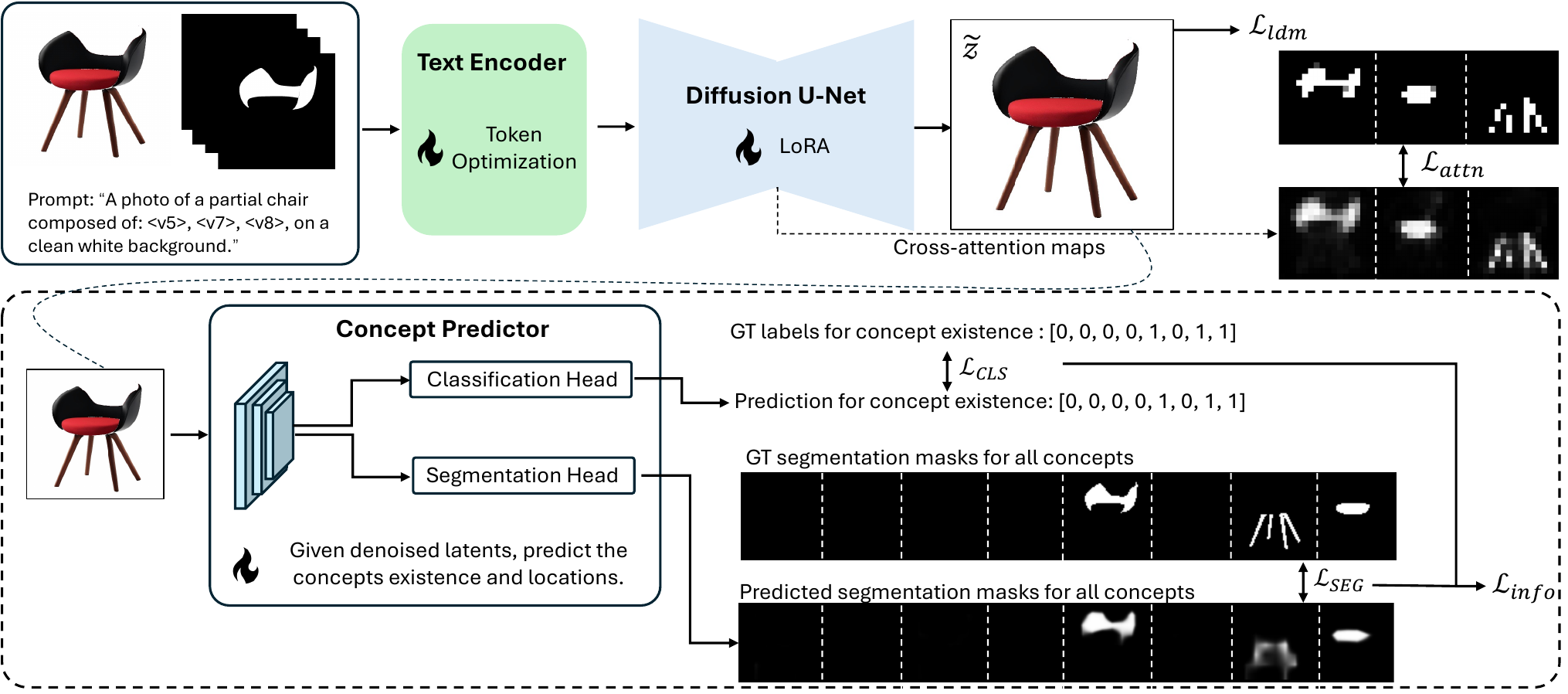}
    \caption{Overview of our method. Given a concept-compositional prompt and noise input, the denoising U-Net produces a latent $\tilde{z}$ supervised by reconstruction loss $\mathcal{L_{\text{ldm}}}$, cross-attention loss $\mathcal{L_{\text{attn}}}$, and information loss $\mathcal{L_{\text{info}}}$. $\mathcal{L_{\text{info}}}$ is computed by a concept predictor which receives $\tilde{z}$ and outputs concept classification and segmentation predictions. The goal is to maximize mutual information between the denoised latent features and the intended concept codes. All modules are jointly optimized to enable part-level concept disentanglement and controllable composition without concept missing or ambiguity.}
    \label{fig:method-pipeline}
\end{figure*}

\paragraph{Dynamic Data Synthesis}

We propose a dynamic data synthesis approach to augment the limited single-image example input. Our goal is to explore and augment the part combination variety while preserving structural integrity information in the diffusion prior. We formalize the training batch as two images, an instance image that is directly sampled from the given examples, and a synthetic image that is generated on-the-fly. Fig.~\ref{fig:method-data-synth} demonstrates our dynamic data synthesis approach. To preserve the structural information of the diffusion prior, we randomly choose an image from the input examples as the instance image, since the input examples already contain reasonable object structures. We randomly select a subset of the parts contained in the chosen instance image, inspired by the union sampling method in Break-a-Scene~\cite{23_Break-a-Scene}. We mask out the unused parts areas and change the background to white to focus only on the part concepts, and pair it with a descriptive prompt. However, this data alone cannot provide enough training data to fine-tune diffusion models to compose different compositions of parts across different images. Thus, to explore and augment part combination variety, we propose to synthesize an image that contains parts that are randomly sampled across input examples. For each part category (e.g., the armrest of the chair), we randomly select a part from the input instances. Inspired by MuDI~\cite{24_MuDI}, we randomly scale and place the sampled parts on a white image. The parts can overlap with each other to encourage learning concepts from multiple possible compositions. We modify the original masks according to the overlapping occlusions. Due to the high variability in the synthetic image, we use a different prompt to treat it as a collection of parts. 

\paragraph{Maximizing Mutual Information}

When observing the poor performance of existing concept learning methods on learning part-level concepts, we argue that a key missing point is that there are no explicit regulations on the information encoded in the concept embeddings and the composition process of the generative model. We thus propose a scheme to maximize the mutual information between the denoised latents and the concepts contained in the input image, and thus disentangle the information encoded in each part-level concept and supervise the diffusion generative process. 

To this end, we adopt a mutual information maximization objective inspired by InfoGAN~\cite{16_InfoGAN}, which encourages the learned concept embeddings to retain interpretable and disentangled semantic structure. Let $\tilde{z}$ denote the denoised latent representation which is generated by a single denoising step of the diffusion model, and let $\mathbf{c}$ denote the set of concept codes associated with the input image. We seek to maximize the mutual information $I(\mathbf{c}; \tilde{z})$, which quantifies how much information about the concepts is preserved in the latent. This can be expressed as:
\[
I(\mathbf{c}; \tilde{z}) = H(\mathbf{c}) - H(\mathbf{c} \mid \tilde{z}),
\]
where $H(\mathbf{c})$ is the entropy of the concept distribution and $H(\mathbf{c} \mid \tilde{z})$ is the conditional entropy of the concepts given the latent. Since directly computing this term is intractable, we introduce a variational distribution $Q(\mathbf{c} \mid \tilde{z})$, implemented via a neural network - concept predictor, to approximate the true posterior. This leads to a variational lower bound:
\[
\mathcal{I}_{\text{lower}} = \mathbb{E}_{\mathbf{c} \sim P(\mathbf{c}), \tilde{z} \sim P(\tilde{z} \mid \mathbf{c})} [\log Q(\mathbf{c} \mid \tilde{z})] + H(\mathbf{c}),
\]
which satisfies $\mathcal{I}_{\text{lower}} \leq I(\mathbf{c}; \tilde{z})$. For a fixed prior over concepts (i.e., the ground-truth labels that supervise the concept predictor), $H(\mathbf{c})$ is constant and can be omitted during optimization. The corresponding training loss $\mathcal{L}_{\text{Info}}$ is defined as the negative of the lower bound
:
\[
\mathcal{L}_{\text{Info}} = - \mathbb{E}_{\mathbf{c} \sim P(\mathbf{c}), \tilde{z} \sim P(\tilde{z} \mid \mathbf{c})} [\log Q(\mathbf{c} \mid \tilde{z})],
\]
which is minimized alongside the overall training loss. This encourages the model to produce latents $\tilde{z}$ from which the concept codes $\mathbf{c}$ can be accurately inferred, effectively regularizing the concept embedding space to reflect correct part-level visual information. This framework provides direct supervision to disentangle part-level concepts and penalizes the generative models from generating ambiguous or missing part combinations.

To effectively implement a $Q(\mathbf{c} \mid \tilde{z})$, we propose the concept predictor design as shown in Fig.~\ref{fig:method-pipeline}, which includes two output heads to provide both classification and segmentation predictions of concepts given a denoised latent. The classification loss $\mathcal{L_{\text{CLS}}}$ penalizes the wrong compositions of concepts (i.e., missing concepts or containing more concepts). The segmentation loss $\mathcal{L_{\text{SEG}}}$ further penalizes wrong localization of concepts (i.e., wrong or entangled location of concepts). A visualized example of the ground truth localization masks and predicted localization segmentation is shown in Fig.~\ref{fig:method-pipeline}. We also provide visualization for the convergence of the segmentation head throughout the training steps in Section~\ref{subsec:vis_concept_predictor}. These two losses are weighted to have the same scale and combined to get the mutual information loss $\mathcal{L}_{\text{Info}}$. The concept predictor is jointly optimized with the concept learning process. 

\paragraph{Overall pipeline}
Our overall training loss consists of the latent diffusion loss $\mathcal{L}_{\text{ldm}}$ and cross-attention loss $\mathcal{L}_{\text{attn}}$ from the original Break-a-Scene pipeline, with our concept prediction loss composed of the classification loss $\mathcal{L}_{\text{CLS}}$ and segmentation loss $\mathcal{L}_{\text{SEG}}$. We also use an auxiliary background loss $\mathcal{L}_{\text{BG}}$ to further improve the generated image quality, which penalizes the generation of content outside the union of selected part masks. This encourages the model to focus on concept-relevant regions and avoid unrelated artifacts in unmasked areas. 
Thus, the total loss is defined as:
$
\mathcal{L}_{\text{total}} = \mathcal{L}_{\text{ldm}} + \mathcal{L}_{\text{attn}} + \lambda_{\text{info}} ( \mathcal{L}_{\text{CLS}} + \lambda_{\text{seg}} \mathcal{L}_{\text{SEG}}) + \lambda_{\text{bg}} \mathcal{L}_{\text{BG}},
$
with the following loss weights:
$\lambda_{\text{info}} = 0.05$, $\lambda_{\text{seg}} = 10.0$, $\lambda_{\text{bg}} = 0.01$.
\section{Experiments}

In this section, we first introduce our experiment setup in terms of data and mask generation, implementation details, and comparison details. We then show the qualitative results and comparisons, followed by quantitative evaluations.

\subsection{Experiments Setup}

\paragraph{Data and Mask Generation} We use both synthetic data and realistic images in our experiments. We generate the synthetic images using DeepFloyd IF~\cite{Deepfloyd_IF}, and we collect realistic images mainly from renderings of 3D objects~\cite{RenderHUb}. 

Our method assumes access to part-level masks for training, but remains agnostic to how these masks are obtained. In practice, we adopt one of the following strategies depending on the dataset:

\begin{itemize}
    \item \textbf{Automatic segmentation and labeling:} We apply off-the-shelf segmentation models such as SAM~\cite{23_SAM} to firstly produce over-segmented masks, followed by GPT-4o-based~\cite{23_GPT_4o} captioning and labeling to group and assign part identities.
    \item \textbf{Manual or direct annotation:} We directly specify or provide part masks, either manually segmenting the parts or from existing annotated assets.
\end{itemize}

Our framework is compatible with any source of part-level supervision, including automatic and manual pipelines. Since the main focus of our work is not on the mask generation itself, we treat it as a pre-process step and do not overemphasize our effort on it.

\paragraph{Implementation Details} We use Stable Diffusion v2.1~\cite{Stable-Diffusion} as the pre-trained text-to-image diffusion model and apply LoRA~\cite{22_LoRA} with rank 32 to the U-Net module. The concept predictor is a CNN network that operates on denoised latents and outputs both classification and segmentation predictions. It consists of three convolution layers (with output channels 16, 32, and 64), followed by two parallel output heads: a classification head composed of two fully connected layers for predicting multi-label concept presence, and a segmentation head consisting of a $1 \times 1$ convolution followed by bilinear upsampling to produce per-concept spatial masks. We follow a two-stage training scheme similar to Break-a-Scene~\cite{23_Break-a-Scene}, where we only update the token embeddings with a high learning rate ($10^{-4}$) in the first stage and fully update both text encoder and LoRA weights in the U-Net with a low learning rate ($10^{-6}$) in the second stage. More training and inference details are introduced in Supplemental A.

\paragraph{Comparison Details}
For finetune-based approaches in concept learning, we compare our methods with 3 representative methods: Break-a-Scene~\cite{23_Break-a-Scene}, which is representative for concept learning from a single image, PartCraft~\cite{23_PartCraft}, which is representative for part-level concept learning from large datasets, and MuDI~\cite{24_MuDI}, which is representative for disentangling subject-level concepts. To align their setting with our task, we adapt the original Break-a-Scene input dataset into a single-image examples input manner and feed part-decomposed examples into the PartCraft and MuDI pipeline. We also conduct comparisons with PiT~\cite{25_PiT}, which trains dedicated priors for different categories and encodes and recomposes part-level concepts at test time. We directly use their provided checkpoints and prepare corresponding examples to conduct inference and compare with our results. More details are explained in Supplemental B.

\subsection{Results}

\subsubsection{Qualitative Evaluation}
We provide qualitative evaluation and comparisons of our pipeline with other methods. We use two tasks - intra-category and cross-category inputs, each with more than 5 distinctive examples, to demonstrate the effectiveness of our pipeline in part-level concept learning and composing.

\paragraph{Intra-category Results}

\begin{figure*}[htbp]
    \centering
    \includegraphics[width=\linewidth]{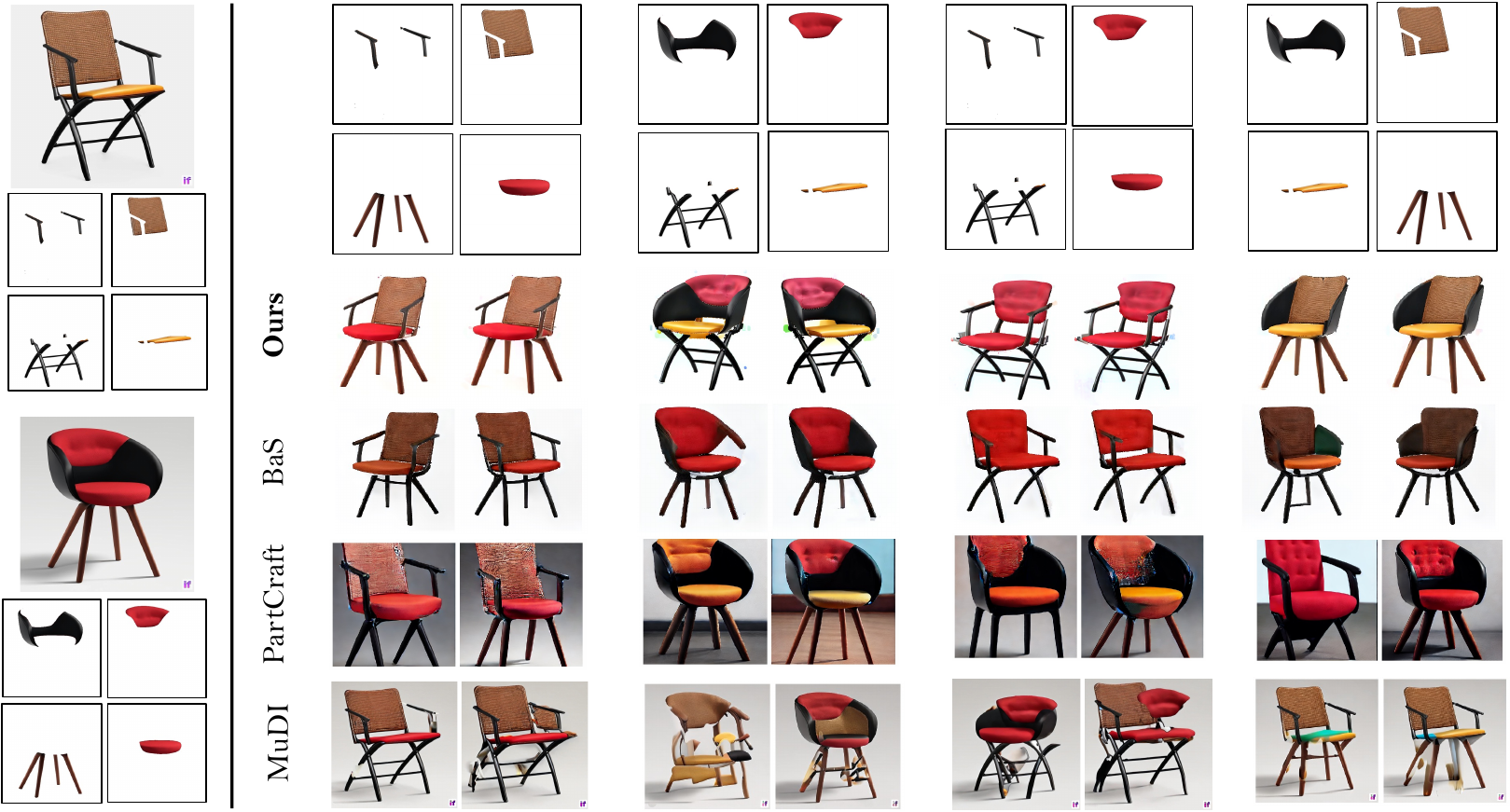}
    \caption{Comparison of concept composition results for 2 chairs using our approach, Break-a-Scene (BaS)~\cite{23_Break-a-Scene}, PartCraft~\cite{23_PartCraft}, and MuDI~\cite{24_MuDI}. The input images and corresponding part decompositions are shown in the left column. We illustrate 4 part compositions across the two input images by selecting 2 parts from each image, which are shown on the top row. We show 2 random samples for each composition.}
    \label{fig:results-intra-chair_comparison}
\end{figure*}

We first demonstrate the part-level concept learning and composition capability of our model using inputs from the same categories. The input subjects consist same part decompositions (e.g., we decompose a chair into 4 parts: armrest, seat back, legs, and seat in Fig.~\ref{fig:results-intra-chair_comparison}.) Note that we do not require the part decompositions to be semantically meaningful since our method can handle any part-level visual concepts.

Fig.~\ref{fig:results-intra-chair_comparison} shows the qualitative comparison of our method with Break-a-Scene~\cite{23_Break-a-Scene}, PartCraft~\cite{23_PartCraft}, and MuDI~\cite{24_MuDI}, where we aim to recompose concepts across two input images. We generate 4 samples for each part composition. Break-a-Scene~\cite{23_Break-a-Scene} struggles to disentangle and recompose part-level concepts, resulting in mixed identity in different parts, presumably since it was designed to do subject-level concept learning. PartCraft struggles to learn and recompose part-level concepts from single-image examples, resulting in poor and inconsistent image generation quality, presumably since it was designed to train on a large dataset. MuDI struggles to learn effective part-level concepts and results in poor composition capability, presumably since it was designed for a few subject-level concepts disentanglement. Our pipeline produces clear part-level concept disentanglement and clean composition capability.

\begin{figure*}[htbp]
    \centering
    \includegraphics[width=\linewidth]{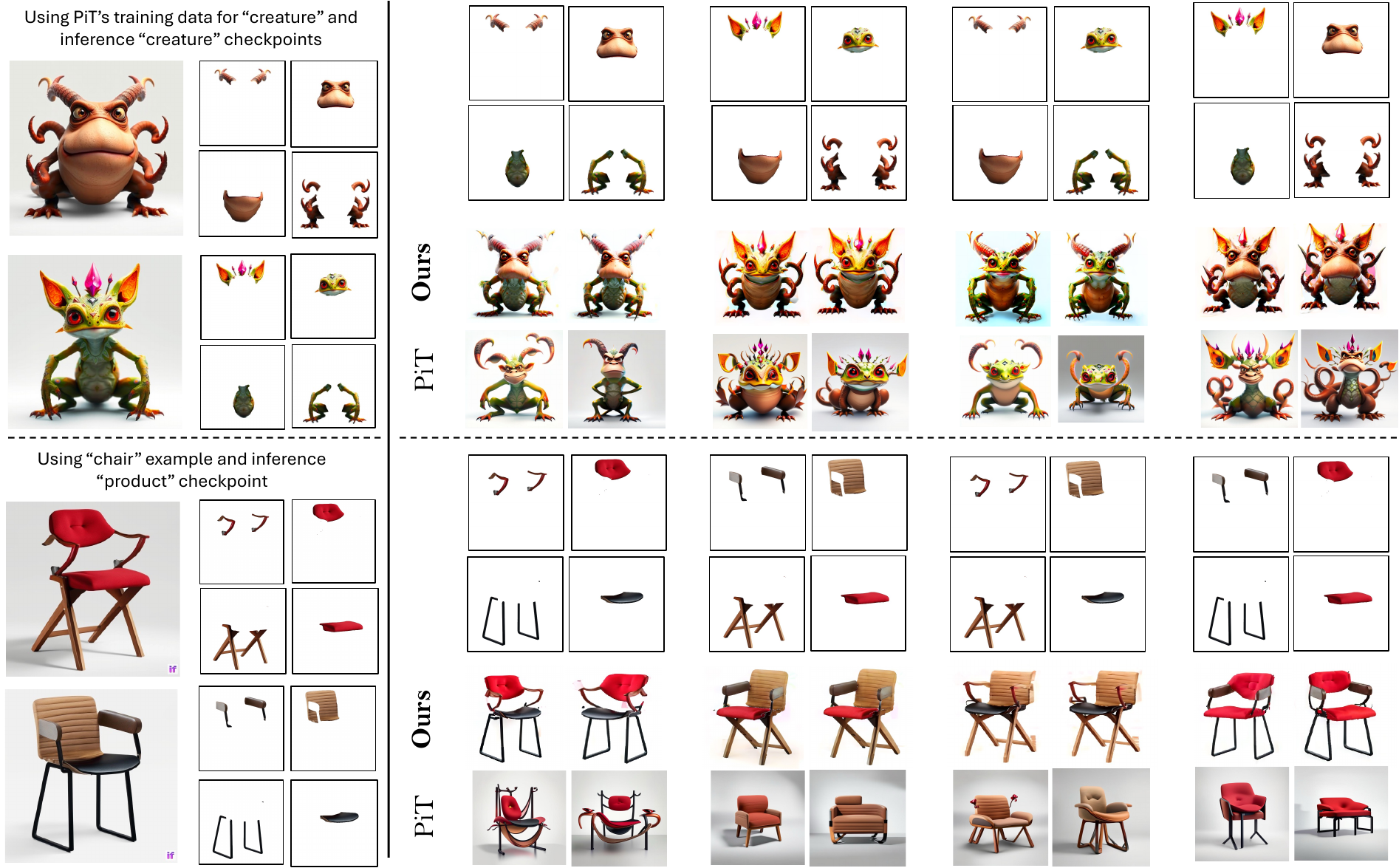}
    \caption{Comparison of concept composition results for 2 creatures and 2 chairs using our approach and PiT~\cite{25_PiT}.}
    \label{fig:results-intra-creature_comparison}
\end{figure*}

To enable fair comparison with PiT~\cite{25_PiT} on their released checkpoints, we conduct two experiments, one for in-distribution and one for out-of-distribution. We first use their training data for the "creature" category and compare our results with the results generated by PiT's "creature" checkpoint. We then use a chair example to compare our results with the results generated by their "product" checkpoint, which is the closest related category in their provided checkpoints. Fig.~\ref{fig:results-intra-creature_comparison} illustrates the qualitative comparison results. Even for in-distribution examples (the "creature", which is directly from their training data), PiT may drastically alter the identity of some concepts, while our method generally preserves decent concept identity. When using out-of-distribution data like chairs, their pipeline fails to encode and compose the provided parts.

 Fig.~\ref{fig:results-pit-v2} illustrates the results for 2 different levels of part granularity. It demonstrates that our method is agnostic to the granularity of parts and can effectively learn a large amount of (more than 4 parts per image) very fine-grained concepts. We also provide substantial results for over 10 other categories in Supplemental C.1.

\paragraph{Cross-category Results}

To enable more creativity in part-level concepts composition, we evaluate our methods in learning and composing parts from objects in different categories, aiming to generate virtual objects. The first row in Fig.~\ref{fig:results-intra-chair_comparison} shows hybrid compositions from a wheelchair and a bed, and the second row in Fig.~\ref{fig:results-intra-chair_comparison} shows hybrid compositions from a wheelchair and a chair. Our methods can generate creative virtual objects with different part compositions, preserving clear part-level identity with reasonable structural arrangements for most composition scenarios. More results are shown in Fig.~\ref{fig:teaser} and in Supplemental C.2.

\subsubsection{Quantitative Evaluation}
We quantitatively compare our pipeline with baselines in terms of concept preserving capability and overall image generation quality. We run experiments for 3 categories (chair, vehicle, and characters), each with 7 single-image examples, and take the averaged score. For a single-image example containing $m$ images and with $k$ part decomposition, we sample 36 images for each of the $m^k$ possible part combinations, resulting in a total of around $1$k image samples per example.

\paragraph{Concept Preserving}

Our goal is to evaluate whether the intended part-level concepts are preserved in the generated images. Several previous works~\cite{23_Break-a-Scene, 24_MuDI} that focus on subject-level concept learning use detectors to locate the subject and then compare feature-wise distance (e.g., DINO~\cite{DINO}) on the located areas. However, automatically detecting fine-grained parts from subjects is hard, and DINO feature comparison is not effective on parts that are small or thin. Thus, we follow the evaluation methods used in PiT~\cite{25_PiT}, which use a multi-modal LLM - QWen~\cite{QWen3} to compare the generated images with intended parts and score the visual alignment. Given $k$ parts to compose a subject, we prompt QWen to score from $0$ to $k$, indicating how many parts are preserved in a generated image. We then normalize the scores for all of the examples into a 0-5 scale and report in Table~\ref{tab:quali_concept}. Our method significantly outperforms baselines to preserve the most intended part-level concepts.

\begin{table}[!htbp]
\caption{Concept preserving comparison. Higher is better ($\uparrow$).}
\centering
\resizebox{0.9\columnwidth}{!}{%
\begin{tabular}{l|ccccc}
\toprule
\textbf{Category} & \textbf{BaS} & \textbf{PartCraft} & \textbf{MuDI} & \textbf{PiT} & \textbf{PartComposer} \\
\midrule
Chairs     & 3.66 & 1.89 & 1.97 & 2.70 & \textbf{4.91} \\
Characters & 4.10 & 3.20 & 2.70 & 4.30 & \textbf{4.89} \\
Vehicles   & 3.07 & 1.85 & 2.03 & 2.57 & \textbf{4.68} \\
\bottomrule
\end{tabular}
}
\label{tab:quali_concept}
\end{table}

\paragraph{Generation Quality.} We evaluate the overall image quality for the generated images, using 2 metrics to evaluate both the overall samples' quality and single image comparison quality. We use FID and KID scores to evaluate the image quality by comparing the sampled images for all possible part combinations and reference image datasets for the corresponding category. For KID, we randomly divide the samples into 20 batches. The reference images are generated using SDXL~\cite{SDXL}. We also use multi-model LLMs to evaluate the generated image quality by directly comparing the samples with the given example images. We use QWen to score the quality on a 1-5 scale on the 5 methods for each example, where the best quality method gets a score of 5 and the worst quality method gets a score of 1. Table~\ref{tab:quali_quality} shows the results. Our model generally achieves the best image quality when compared to other models.

\begin{table}[!htbp]
\caption{Generation quality comparison for 3 categories.}
\centering
\resizebox{\columnwidth}{!}{%
\begin{tabular}{ll|ccccc}
\toprule
\textbf{Category} & \textbf{Metrics}       & \textbf{BaS} & \textbf{PartCraft} & \textbf{MuDI} & \textbf{PiT} & \textbf{PartComposer} \\
\midrule
\multirow{4}{*}{Chair} 
         & FID $\downarrow$           & 17.22 & 26.91 & 16.07 & 23.95 & \textbf{15.85} \\
         & KID - Mean $\downarrow$    & 0.0632 & 0.1098 & 0.0551 & 0.0800 & \textbf{0.0502} \\
         & KID - Std Dev $\downarrow$ & 0.0066 & 0.0097 & \textbf{0.0064} & 0.0085 & 0.0067 \\
         & QWen Score $\uparrow$      & 2.87 & 1.48 & 3.72 & 2.21 & \textbf{4.72} \\
\midrule
\multirow{4}{*}{Vehicle} 
         & FID $\downarrow$           & 20.54 & 18.78 & 22.26 & 20.12 & \textbf{18.30} \\
         & KID - Mean $\downarrow$    & 0.2066 & 0.1751 & 0.2342 & 0.1772 & \textbf{0.1530} \\
         & KID - Std Dev $\downarrow$ & 0.0093 & 0.0092 & 0.0113 & \textbf{0.0092} & 0.0095 \\
         & QWen Score $\uparrow$      & 2.75 & 1.54 & 3.94 & 2.13 & \textbf{4.64} \\
\midrule
\multirow{4}{*}{Creature} 
         & FID $\downarrow$           & 19.26 & 20.87 & 25.06 & 17.56 & \textbf{17.26} \\
         & KID - Mean $\downarrow$    & 0.1314 & 0.1692 & 0.2038 & 0.1179 & \textbf{0.1018} \\
         & KID - Std Dev $\downarrow$ & \textbf{0.0065} & 0.0098 & 0.0071 & 0.0066 & 0.0067 \\
         & QWen Score $\uparrow$      & 3.10 & 1.64 & 1.60 & 3.98 & \textbf{4.68} \\
\bottomrule
\end{tabular}
}
\label{tab:quali_quality}
\end{table}

\subsubsection{Flexibility and Scalability}
We demonstrate the flexibility and scalability of our pipeline in 3 aspects: encoding additional background concepts, dealing with incomplete part combination prompts, and scaling to learn a large number of concepts from more than 2 single-image examples. The left column in Fig.~\ref{fig:flex_bg_partial} shows that our pipeline can learn additional background concepts and naturally blend the composed object in the background. The right column in Fig.~\ref{fig:flex_bg_partial} shows that our method can generate structurally complete objects given incomplete part combinations and can have variety in unspecified parts. The last row in Fig.~\ref{fig:results-fig-addtional-intra-cross} shows that our method can directly learn a large number of concepts (i.e., 16 concepts from 4 single-image examples). All previous works often struggle to learn over 5 concepts at the same time without using large datasets.

\subsubsection{Ablation Study}
\begin{figure}[htbp]
\centering
\begin{tabular}{c@{\hspace{4pt}}c@{\hspace{4pt}}c@{\hspace{4pt}}c}
Compositions & w/o D.S. & w/o C.P & Full approach \\
\includegraphics[width=0.20\columnwidth]{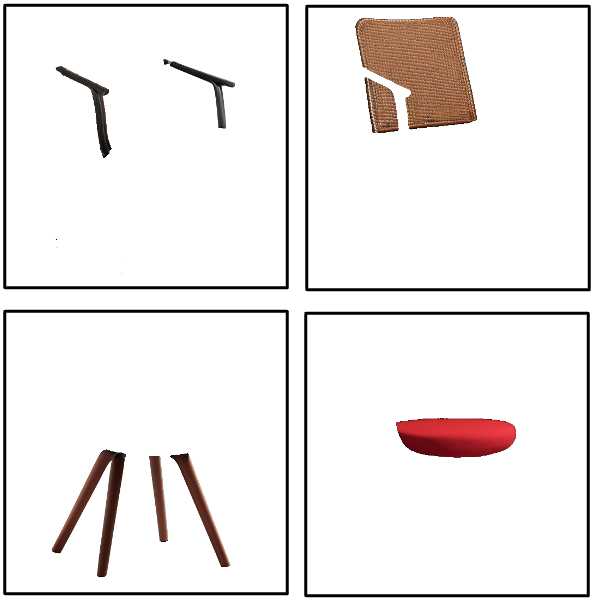} &
\includegraphics[width=0.20\columnwidth]{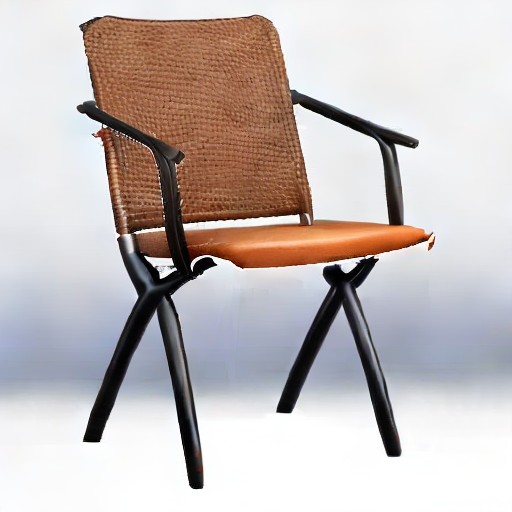} &
\includegraphics[width=0.20\columnwidth]{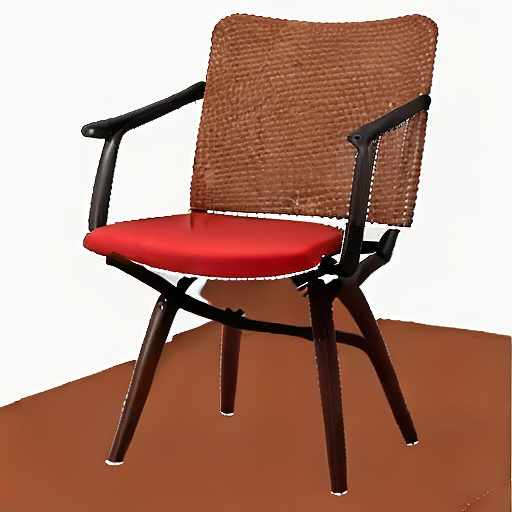} &
\includegraphics[width=0.20\columnwidth]{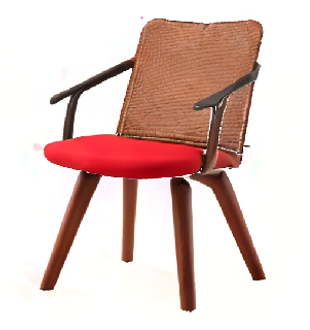} \\
\includegraphics[width=0.20\columnwidth]{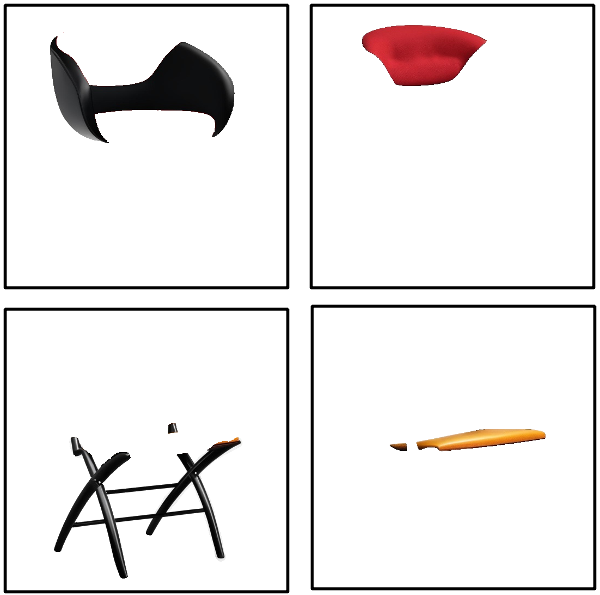} &
\includegraphics[width=0.20\columnwidth]{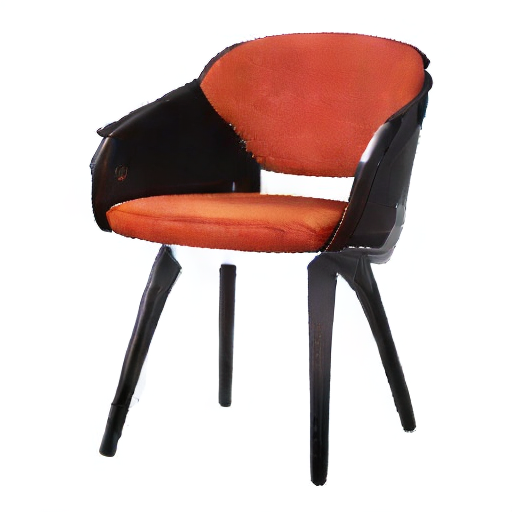} &
\includegraphics[width=0.20\columnwidth]{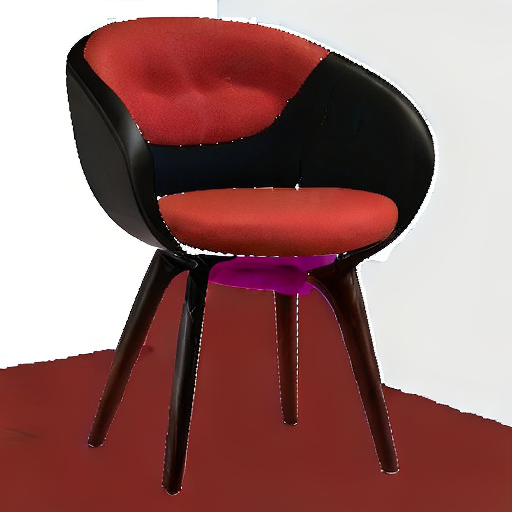} &
\includegraphics[width=0.20\columnwidth]{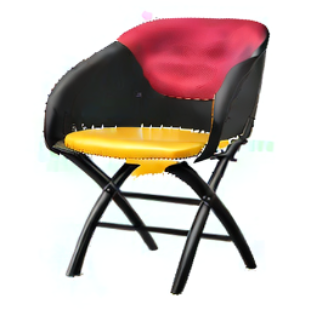} \\
\includegraphics[width=0.20\columnwidth]{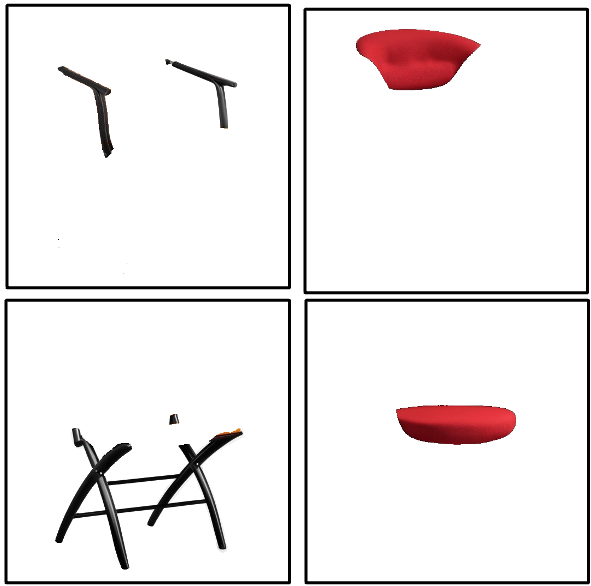} &
\includegraphics[width=0.20\columnwidth]{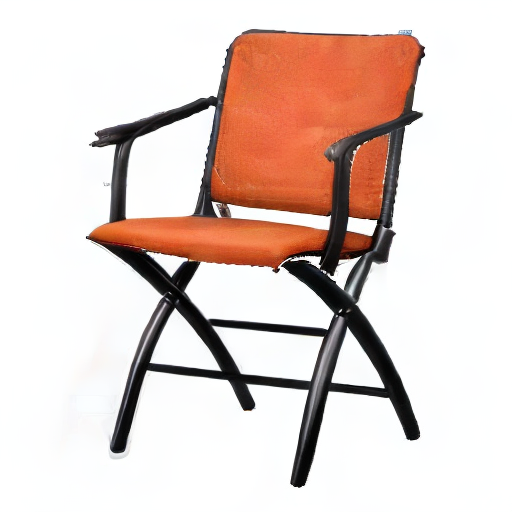} &
\includegraphics[width=0.20\columnwidth]{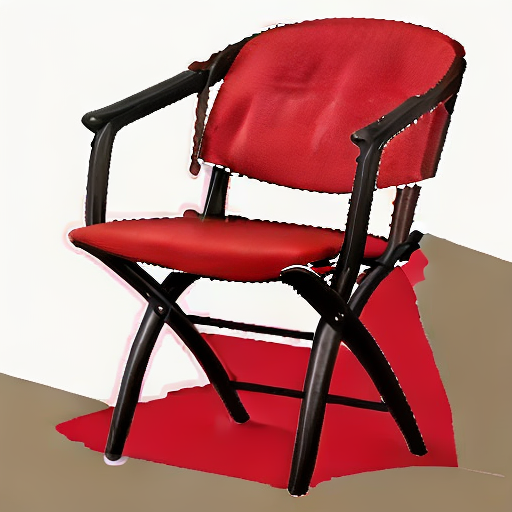} &
\includegraphics[width=0.20\columnwidth]{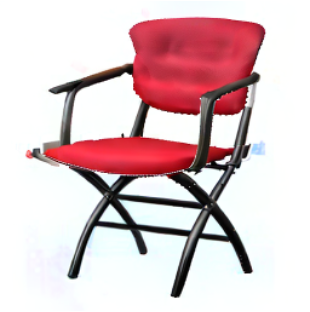} \\
\includegraphics[width=0.20\columnwidth]{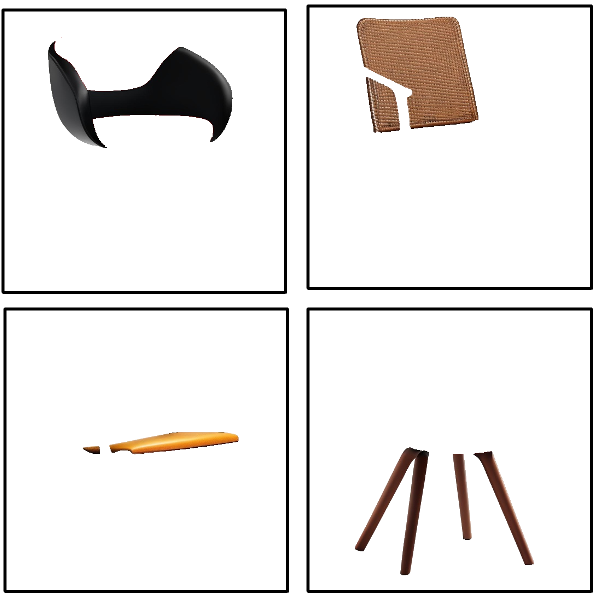} &
\includegraphics[width=0.20\columnwidth]{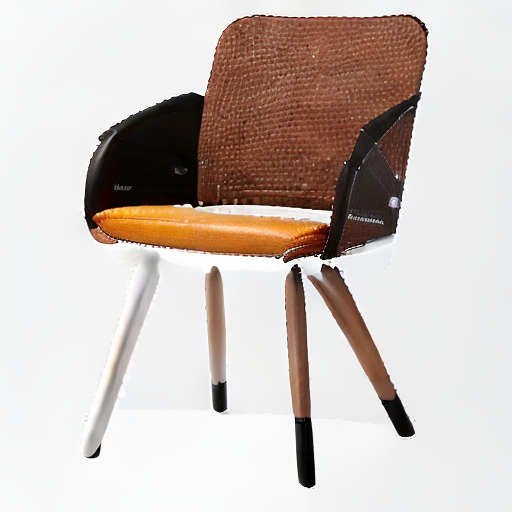} &
\includegraphics[width=0.20\columnwidth]{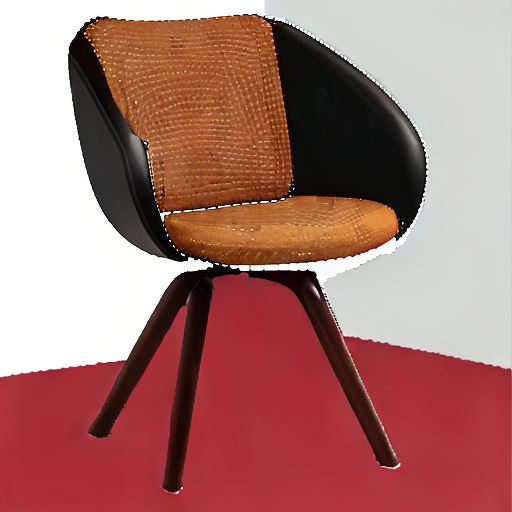} &
\includegraphics[width=0.20\columnwidth]{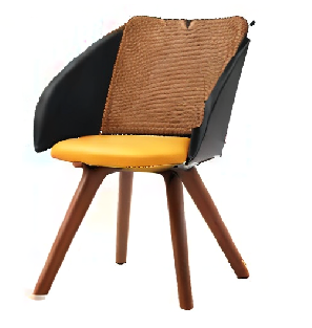} \\
\end{tabular}
\caption{Qualitative comparison across ablations: without dynamic data synthesis (w/o D.S.), without concept predictor (w/o C.P), and our full approach.}
\label{fig:ablation_comparison}
\end{figure}

 We conduct ablation studies by removing the dynamic data synthesis method and the concept predictor. Fig.~\ref{fig:ablation_comparison} illustrates the ablation results. Disabling the dynamic data synthesis results in poor data utility, and disabling the concept predictor results in concept disentanglement or missing.

\subsubsection{Weight and Scale Experiments}
\begin{figure}[htbp]
    \centering
    \includegraphics[width=\linewidth]{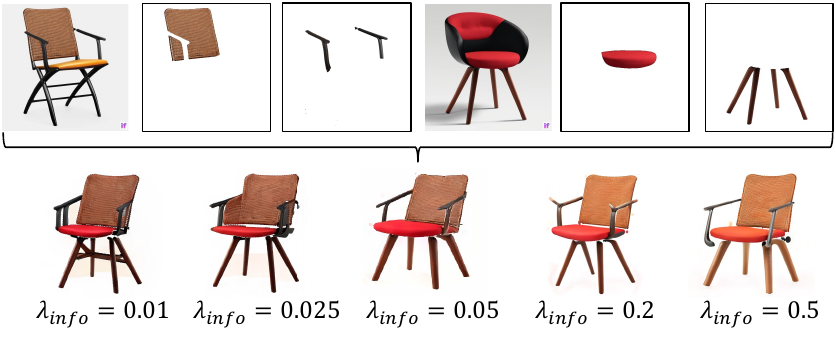}
    \caption{Visualization for the impacts of different information loss weights.}
    \label{fig:analysis-info-weight}
\end{figure}

The most important weight in our experiments is the weight for the concept predictor loss $\lambda_{\text{info}}$ explained in Section~\ref{sec:method}. We found that $\lambda_{\text{info}}=0.05$ is usually the best weight for most examples, though some examples require slightly higher weights ranging from $0.075$ to $0.1$. Increasing the $\lambda_{\text{info}}$ might show obvious improvement on concept preserving but may result in unnatural artifacts since the diffusion prior may get damaged. Fig.~\ref{fig:analysis-info-weight} illustrates this phenomenon. When increasing the weight from $0.01$ to $0.05$, concepts identity preservation is improved, showing the effectiveness of our concept predictor. When further increasing the weight, although concepts are still well-preserved, there are some visual artifacts, including saturation shift.

\subsubsection{Visualization for concept predictor}
\label{subsec:vis_concept_predictor}
\begin{figure}[htbp]
    \centering
    \includegraphics[width=\linewidth]{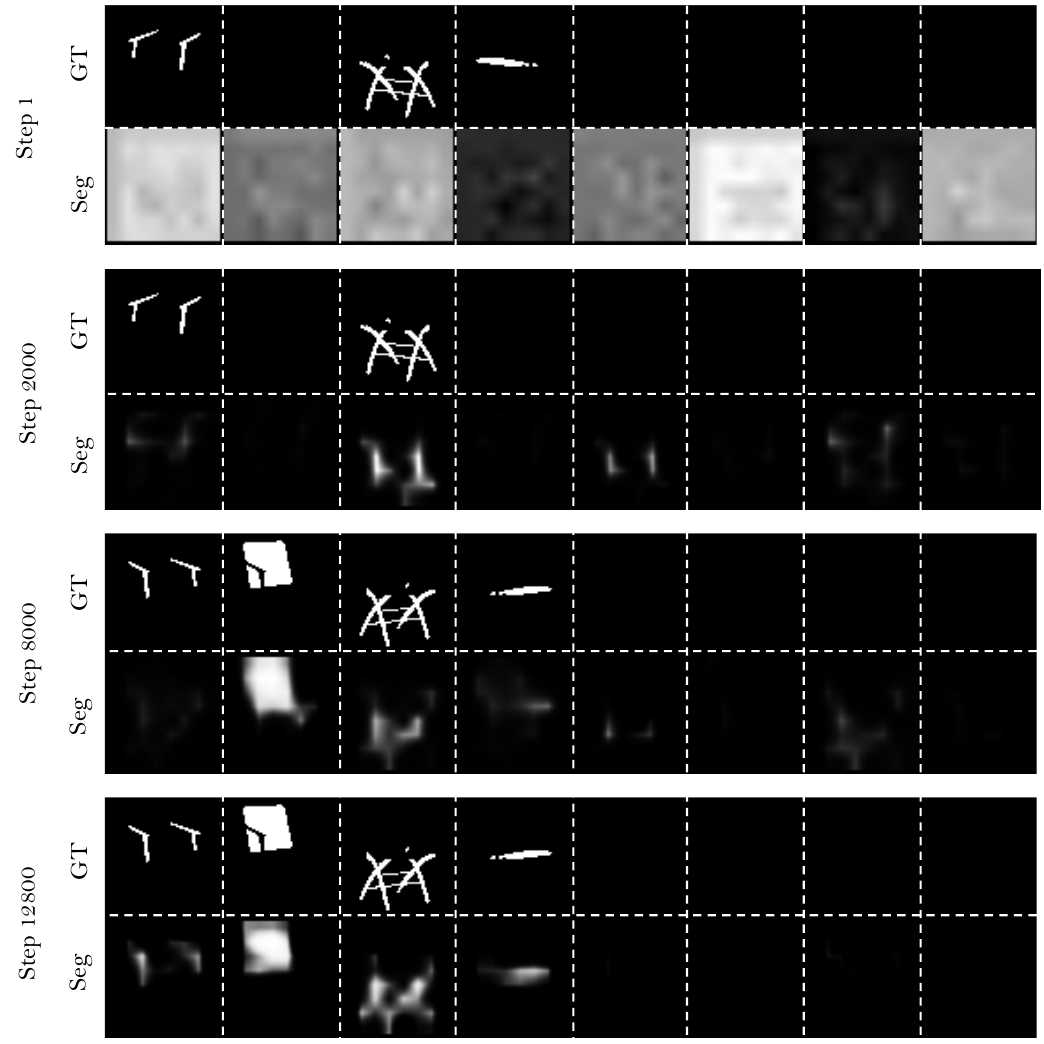}
    \caption{Visualization for the convergence of the segmentation head in our concept predictor.}
    \label{fig:analysis-seg}
\end{figure}

\begin{figure*}[htbp]
    \centering
    \includegraphics[width=0.99\linewidth]{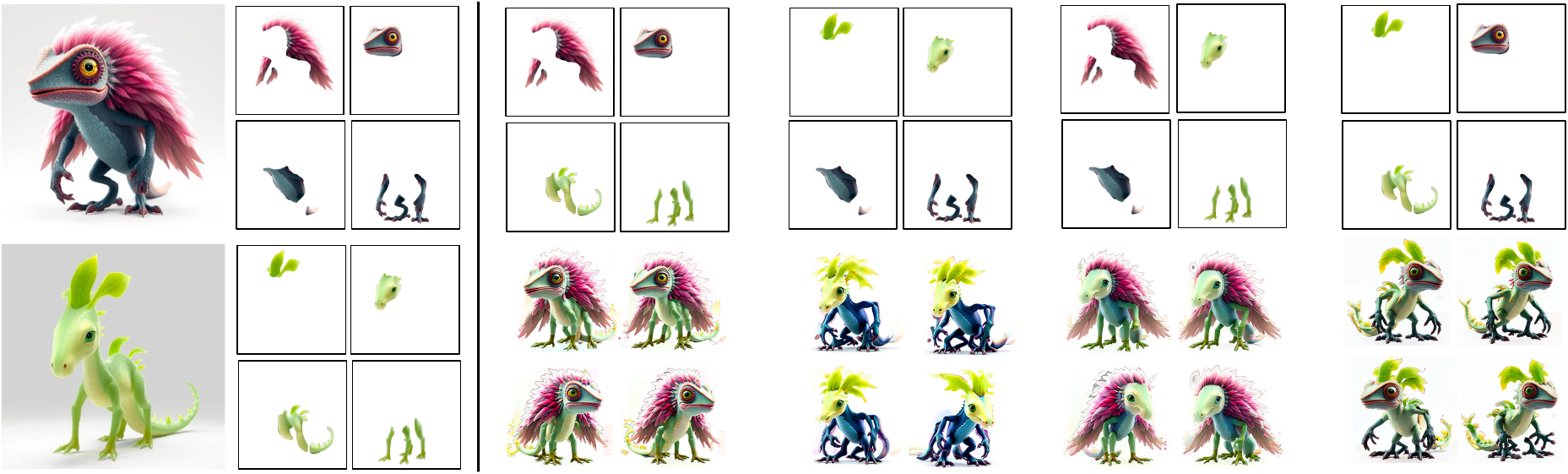}
     \medskip\hrule\medskip
    \includegraphics[width=0.99\linewidth]{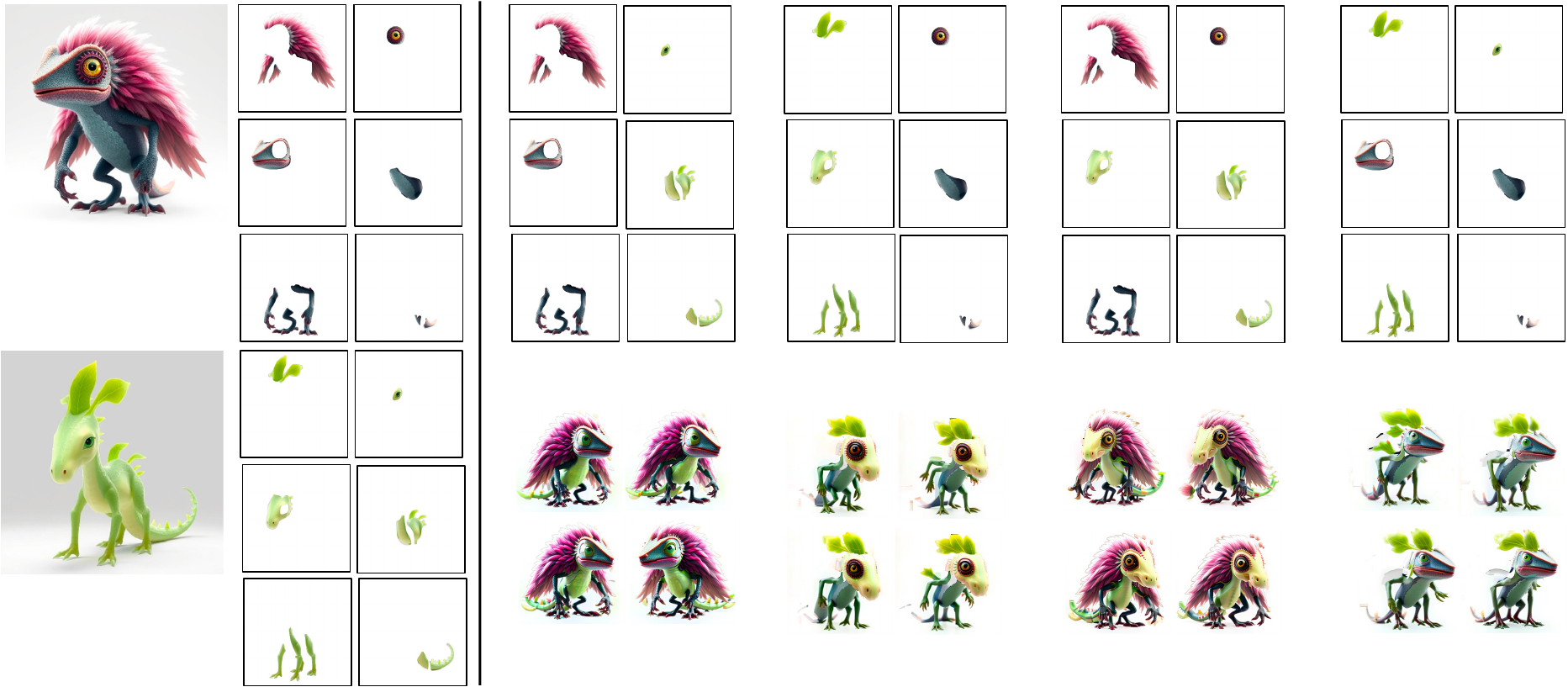}\\
    \caption{Concept learning and composition results for two virtual creatures at coarse and fine part segmentation levels. Our pipeline can effectively learn very fine-grained part-level concepts from single-image examples.}
    \label{fig:results-pit-v2}
\end{figure*}

\begin{figure*}[htbp]
    \centering
    \includegraphics[width=\linewidth]{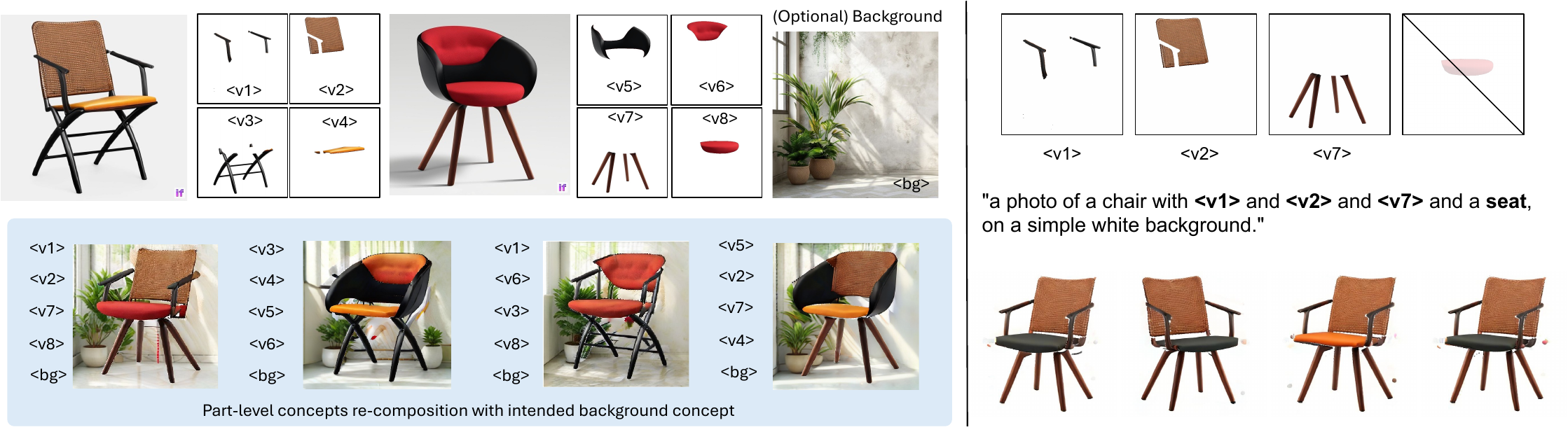}
    \caption{Flexibility of our pipeline: Our pipeline can learn background concept together with part concepts, and naturally blend the composed objects into the background (left column). Our pipeline can also handle incomplete combination of parts in a prompt and generate complete objects with variations in the unspecified parts.}
    \label{fig:flex_bg_partial}
\end{figure*}

\begin{figure*}[htbp]
    \centering
    \includegraphics[width=0.96\linewidth]{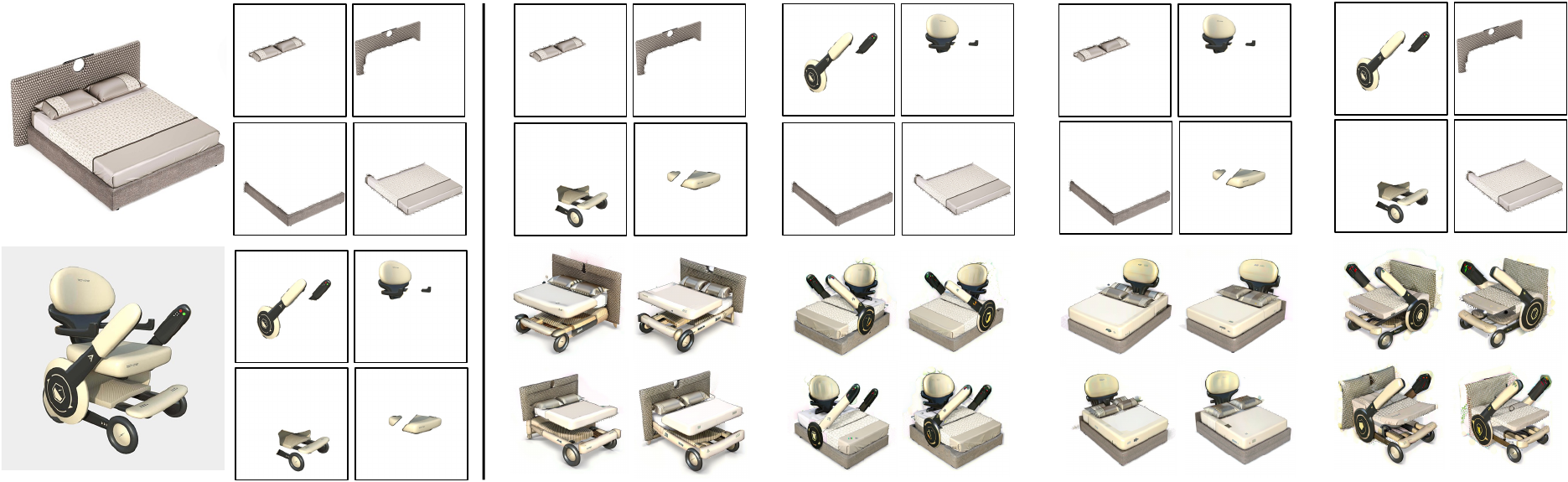}\\
    \medskip\hrule\medskip
    \includegraphics[width=0.96\linewidth]{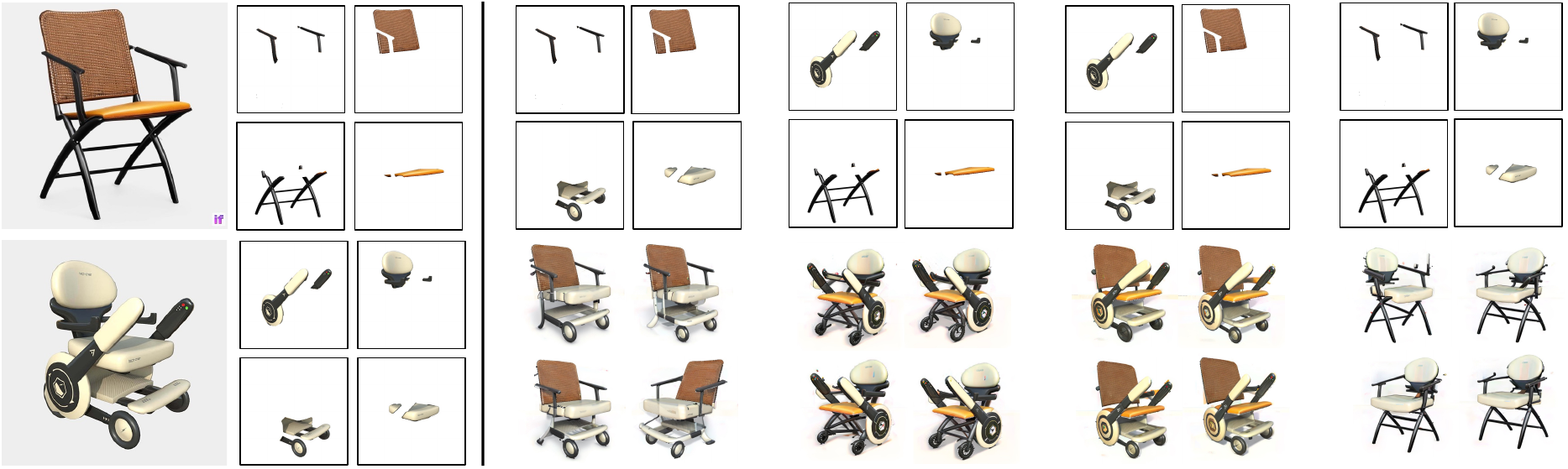}\\
    \medskip\hrule\medskip
    \includegraphics[width=0.96\linewidth]{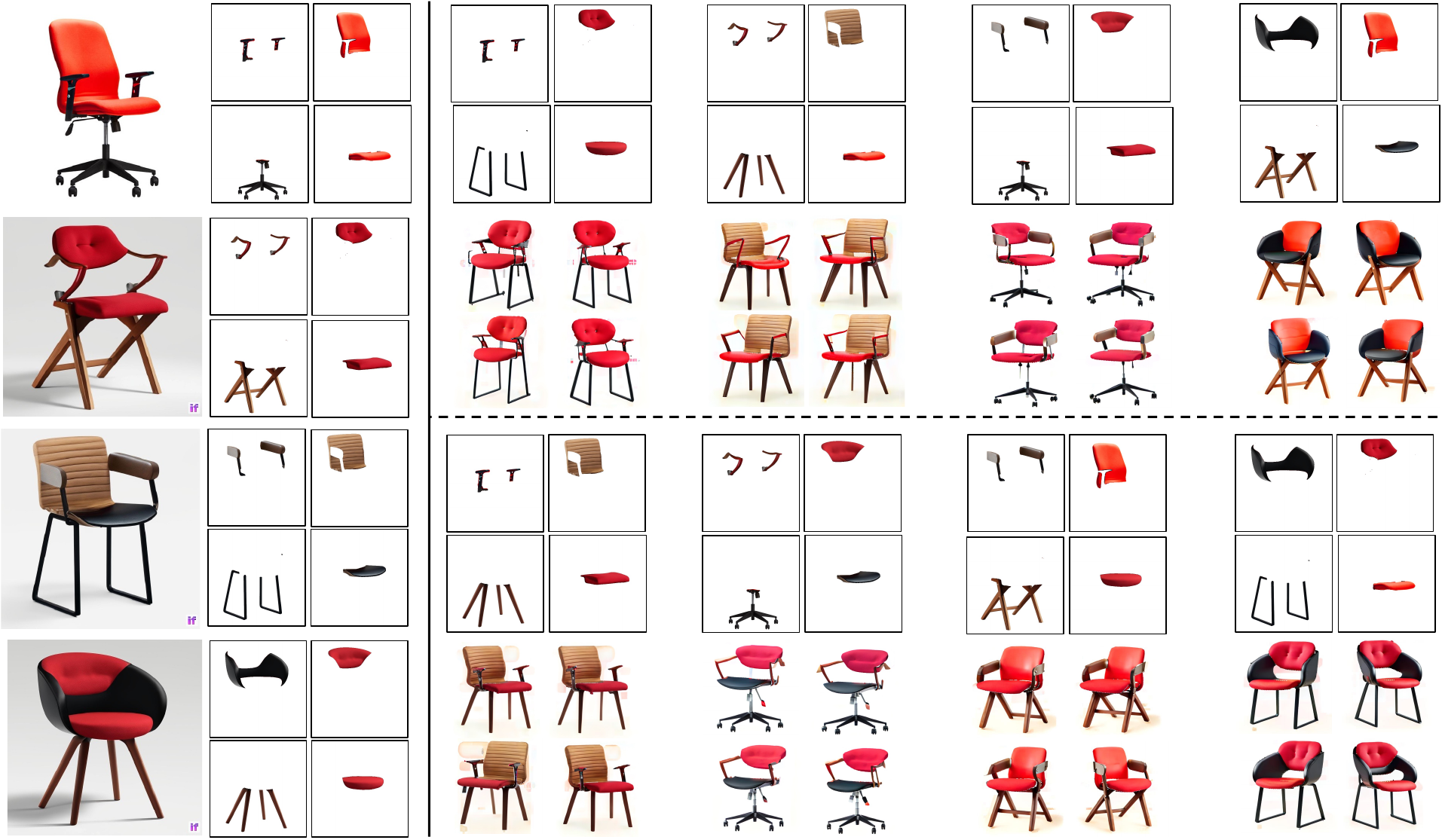}\\
    \caption{Concept learning composition results for more cross-category data and scaled-up single-image examples (16 concepts at one time).}
    \label{fig:results-fig-addtional-intra-cross}
\end{figure*}

We visualize the convergence of our concept predictor's segmentation head to demonstrate the effectiveness of maximizing mutual information. Fig.~\ref{fig:analysis-seg} illustrates the comparison of segmentation results from the concept predictor and the ground truth masks. The concept predictor gradually converges to predict the correct segmentation of the parts, showing the effectiveness of our concept predictor design.

\section{Limitations}
\label{append:limitation}

Although our method can learn well-entangled part-level concepts and re-compose them, the image generation quality for some challenging inputs still need improvement. For example, when composing parts from a truck and a sports car in Supplemental Fig.5, there exist artifacts in the generated images even though all the intended part concepts are preserved. Our pipeline also sometimes does not preserve the exact details for very thin structures, such as the number of bird legs (in Supplemental Fig.9). This problem might be partially due to the inherent limitation of text-to-image diffusion models, since these models can produce images containing unrealistic details. Additionally, for cross-category part re-composition, not all compositions of parts result in meaningful virtual objects. For example, when mixing a chair and a gym equipment in Supplemental Fig.11, the composition in the third column results in non-meaningful objects. A composition prediction scheme may be developed in this case to predict the meaningful part compositions.
\section{Conclusion}

We have introduced PartComposer, a method for fine-grained concept learning that enables new visual generation capabilities through part-level combination and composition. We propose a mutual information maximization framework to enable clear disentanglement of concepts and reduce `missing concept' failure modes. Combined with a dynamic data synthesis procedure, PartComposer is able to learn how to disentangle and compose part-level concepts from single-image examples. Our approach is able to produce natural part compositions which preserve concept identity, from both matching and distinct categories. One promising direction for future work would be to convert our 2D productions into fully realized 3D models, using inverse rendering or image-to-3D pipelines. Looking forward, we believe our method can benefit creative design and imagination workflows, producing novel ideations from easy-to-source inputs, that might help spark new inspiration and creations. 

\begin{acks}
We would like to thank all the anonymous reviewers for their helpful suggestions.  We would like to thank Caitlin Gong and Clara Fee for generating several single-image examples with masks and annotations. We would like to thank all the people whom we had discussions with for thinking about and commenting on interesting and illustrative examples to showcase the part re-composition capability. This work was funded by NSF award \#1941808. Daniel Ritchie is
an advisor to Geopipe and owns equity in the company. Geopipe
is a start-up that is developing 3D technology to build immersive
virtual copies of the real world with applications in various fields,
including games and architecture.
\end{acks}

\bibliographystyle{ACM-Reference-Format}
\bibliography{main}

\newpage

\FloatBarrier  

\newpage
\appendix
\section{Training Details}
\label{append:sec:training and inference}

We provide the detailed training configuration used in our experiments for learning part-level visual concepts from single-image examples. Our pipeline is trained using Stable Diffusion v2.1~\cite{Stable-Diffusion} as the base text-to-image model. We use a pair of \texttt{chair} images from Fig. 3 in the main paper as an illustrative example. Each chair is decomposed into four semantic parts: \texttt{armrest}, \texttt{backrest}, \texttt{legs}, and \texttt{seat}. These parts are tokenized into 8 learnable placeholder tokens, denoted as \texttt{<v1>} through \texttt{<v8>}, with each token corresponding to a specific part instance from the two training images.

\paragraph{Prompt Design.}
Each training image is paired with a structured, part-compositional prompt using the assigned placeholder tokens. Two types of prompts are generated dynamically during training:
\begin{itemize}
    \item \textbf{Instance prompts:} These describe partial objects composed of a subset of parts from a single input image. For example, if parts \texttt{<v5>}, \texttt{<v7>}, and \texttt{<v8>} are selected from Image 2, the corresponding prompt is: \textit{``A photo of a partial chair composed of: \texttt{<v5>}, \texttt{<v7>}, \texttt{<v8>}, on a clean white background.''}
    \item \textbf{Synthetic prompts:} These describe compositions of parts sampled across both input images. For example, if \texttt{<v2>} is sampled from Image 1 and \texttt{<v5>}, \texttt{<v7>}, and \texttt{<v8>} are sampled from Image 2, the prompt is: \textit{``A photo of randomly placed chair components: \texttt{<v2>}, \texttt{<v5>}, \texttt{<v7>}, \texttt{<v8>}, on a clean white background.''}
\end{itemize}
All prompts are automatically generated based on the selected part indices. Backgrounds are set to white by default, and prompts are templated consistently to ensure clean compositional control.

\paragraph{Training Procedure.}
We adopt a two-phase training scheme:
\begin{itemize}
    \item \textbf{Phase 1:} Only concept token embeddings are optimized for 6{,}400 steps.
    \item \textbf{Phase 2:} The LoRA-injected U-Net (with rank 32) and text encoder are jointly fine-tuned for 40{,}000 steps.
\end{itemize}
We want to note that the second stage training step number is just a rough reference, which generally ensures that all concepts are well-learned, and removes background contents. In general, after 18{,}000 steps, the concept learning and re-composition results are already good enough.

\paragraph{Inference Procedure.}
We perform inference using standard DDIM~\cite{20_DDIM} sampling with a pretrained Stable Diffusion v2.1 backbone, the optimized text encoder, and the learned LoRA weights. Given a compositional prompt (e.g., \texttt{"A photo of a partial chair with \texttt{<v2>}, \texttt{<v5>}, \texttt{<v7>}, \texttt{<v8>}, on a clean white background."}), the model decodes a final image using 50 DDIM steps. We use a commonly used guidance scale 7.5.

\paragraph{Learning Extra Background Concepts.}
Our pipeline also supports learning to extract background concepts if the user wants to place the re-composed object on specific background images. We use \texttt{<bgX>} to represent the background concepts, and they are learned together with the part-level concepts. In our data dynamic data synthesis stage, we replace the white background with the given background images and keep all other operations as the same. We use the background loss $\lambda_{\text{bg}}$ with weight $\lambda_{\text{bg}} = 0.01$ to train the background concepts. Fig. 11 in the main paper shows the results for incorporating an indoor background with re-composed chairs. Our method can naturally blend the newly composed chair into the given background.
\section{Comparison Implementation}
\label{append:quali-compare-implement}
We compare our method with other visual concept learning and re-composition works, Break-a-Scene~\cite{23_Break-a-Scene} and PartCraft~\cite{23_PartCraft}. We introduce the detailed comparison experiment setup in this section. 

Break-a-Scene~\cite{23_Break-a-Scene} is originally designed to learn multiple subject-level concepts in a single image. We modify the data loading and processing scheme to support learning part-level concepts from single-image examples. In addition, since we can only access 24G VRAM GPU (RTX3090), we add LoRA~\cite{22_LoRA} following the standard StableDiffusion v2.1 model finetuning scheme to train the LoRA weights instead of the entire diffusion U-Net. This approach has been validated by PartCraft~\cite{23_PartCraft} that it will not harm the concept learning quality when compared to finetuning the entire U-Net. 

PartCraft~\cite{23_PartCraft} is originally designed to learn part-level concepts from a large dataset of images. Specifically, they require to have 10-20 images for a subject (e.g., a specific bird species). We modify the data loading and processing scheme to load the single-image examples dataset into their pipeline. In addition, we directly provide the part-level masks instead of using their automatic concept discovery method to ensure fair comparison. We use StableDiffusion v1.5 as described in their paper to learn the concepts. Their approach shows unstable performance. The results we show in Fig. 5 of the main paper are already the best results we observed. The image quality degrades quickly when the training step number increases (after the step we showed in Fig. 5 of the main paper). We also tried to use a smaller learning rate, but their pipeline struggled to even learn any meaningful concepts. We hypothesize that the reason for this is that their method is designed to be used on a large dataset instead of the single-image examples in our case.

MuDI~\cite{24_MuDI} is originally designed to learn and disentangle similar subject-level concepts from multi-image examples. They generally require over 5 images per subject. We use the SDXL with LoRA as described in their paper to finetune the diffusion model. For the inference, we use their official inference code, which first creates a mean-shifted noise by pasting the intended part onto a blank image, and then adds noise, and then run SDXL~\cite{SDXL} to iteratively denoise. The results we show in Fig. 5 of the main paper are already the best results we observed. Their approach struggles at the part-level and with more than 5 concepts - the single-image example of 2 chairs contains 8 concepts.

PiT~\cite{24_MuDI} is originally designed for creative visual designs by training priors for different object categories and inferring creative designs given different part inputs. This method only requires test-time inference, which is different from all the above methods and our methods, which are optimization-based. Since we don't have access to their full training data, we directly use their released 3 checkpoints (product, creature, and toy) to conduct evaluation. To enable fair comparison, we first use 2 "creature" images that are used in their training and inference through their released "creature" checkpoint. The goal is to evaluate their concept learning and composition capability for in-distribution data. We then use 2 "chair" images and inference through their "product" checkpoint, which is the most related checkpoint in terms of category similarity. The goal for this design is to evaluate their capability to encode and compose parts for slightly out-of-distribution examples.

\section{Extra Experimental Results}

\subsection{Intra-category Results}
\label{append:subsec:intra-category}

We conduct experiments on various object categories to demonstrate our model’s capability in learning and re-composing part-level concepts. In each case, objects are decomposed into over four semantic parts, and our method learns to mix these parts across instances from the same category. We want to note that all the semantic part names are annotated just for ease of reading in the paper. We do not require semantically meaningful part decomposition and detailed annotation for the part-level concepts in our pipeline since we only treat them as visual information.

\begin{figure*}[htbp]
    \centering
    \includegraphics[width=\linewidth]{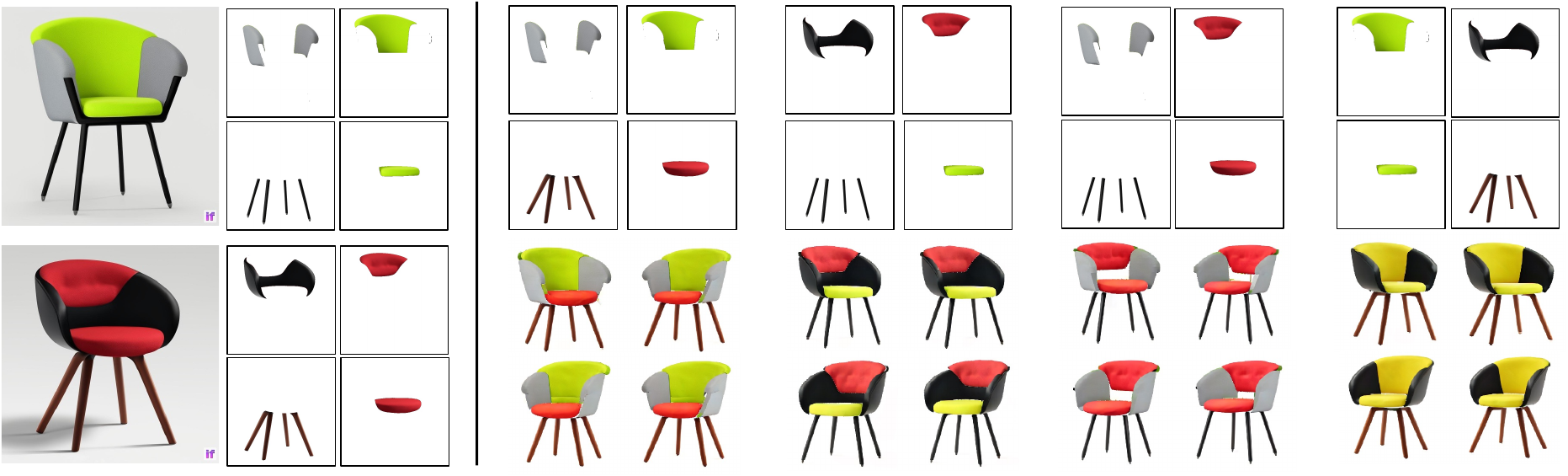} \\
    \hdashrule[0.5ex]{\linewidth}{1pt}{3pt 2pt} \\  
    \includegraphics[width=\linewidth]{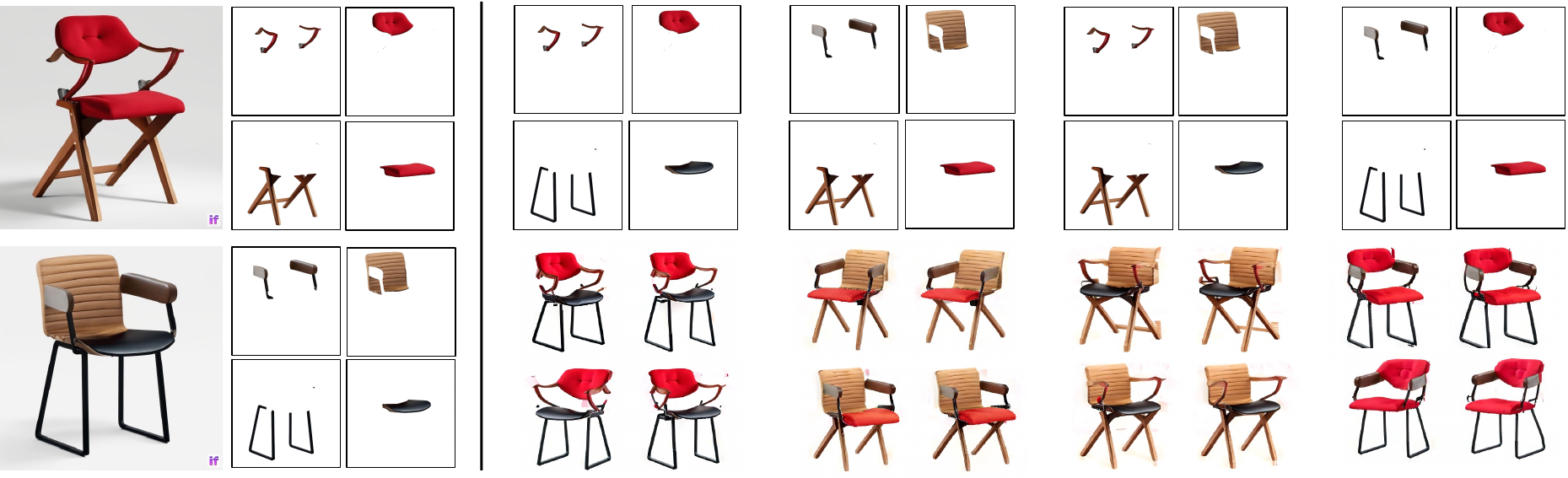} \\
    \hdashrule[0.5ex]{\linewidth}{1pt}{3pt 2pt} \\  
    \includegraphics[width=\linewidth]{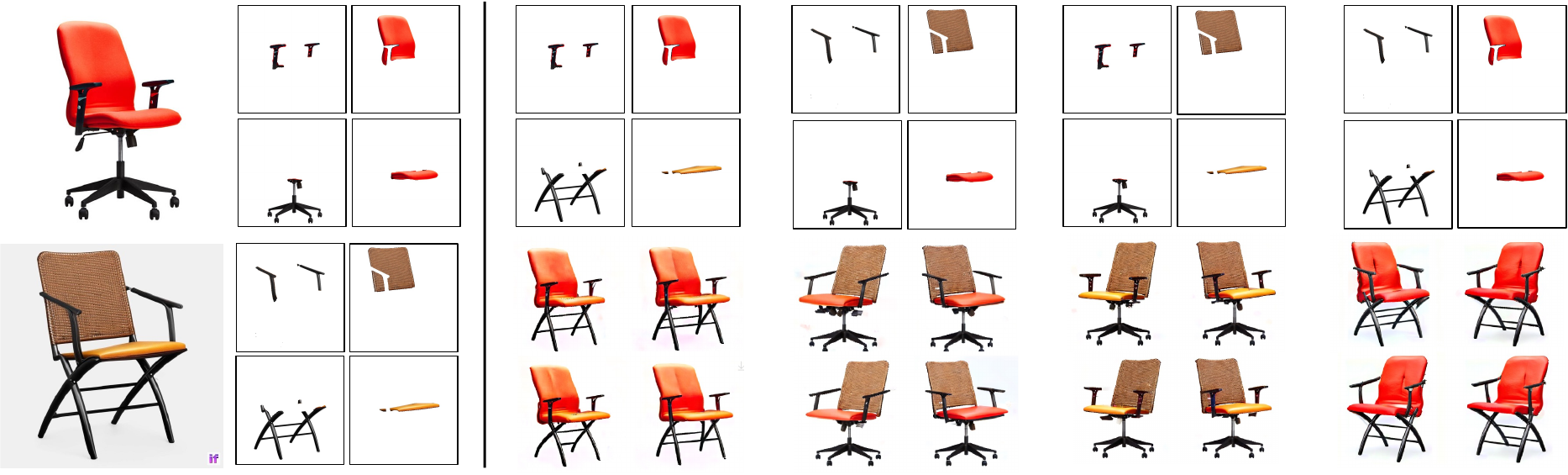}
    \caption{Concept composition results for various compositions of chair parts.}
    \label{fig:results-intra-chairs-1}
\end{figure*}

\begin{figure*}[htbp]
    \centering
    \includegraphics[width=\linewidth]{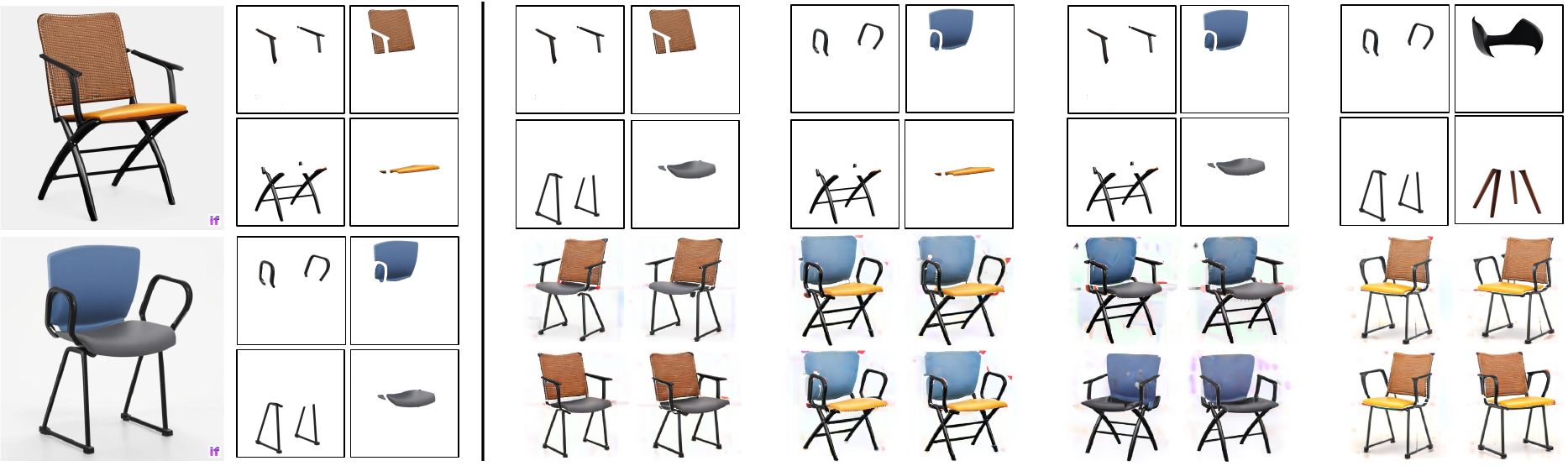} \\
    \hdashrule[0.5ex]{\linewidth}{1pt}{3pt 2pt} \\  
    \includegraphics[width=\linewidth]{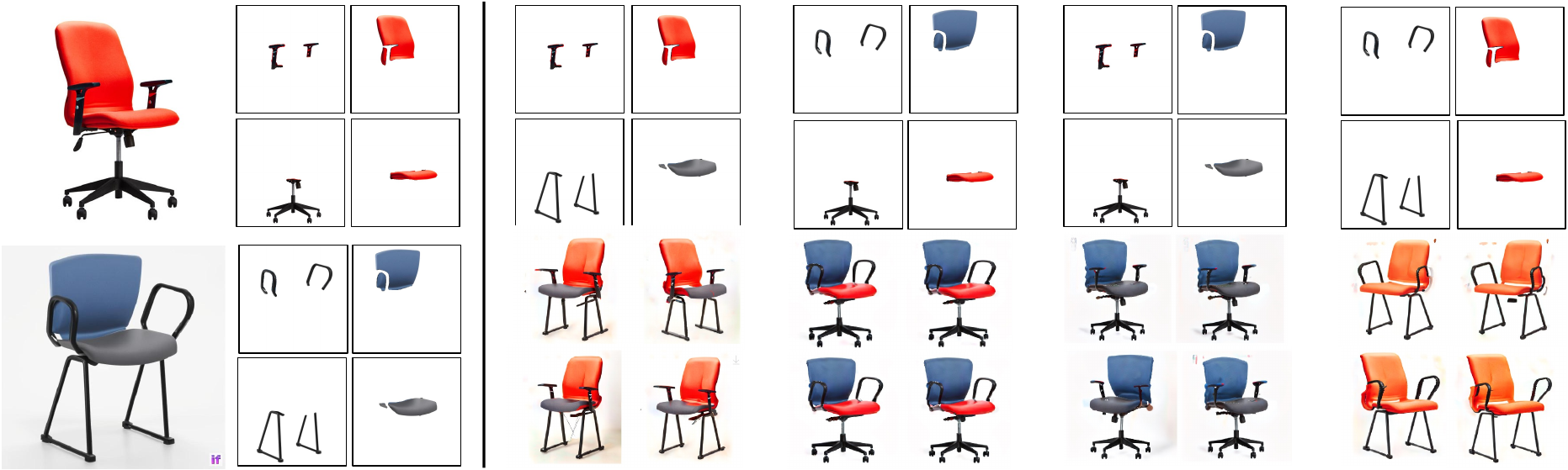} \\
    \hdashrule[0.5ex]{\linewidth}{1pt}{3pt 2pt} \\  
     \includegraphics[width=\linewidth]{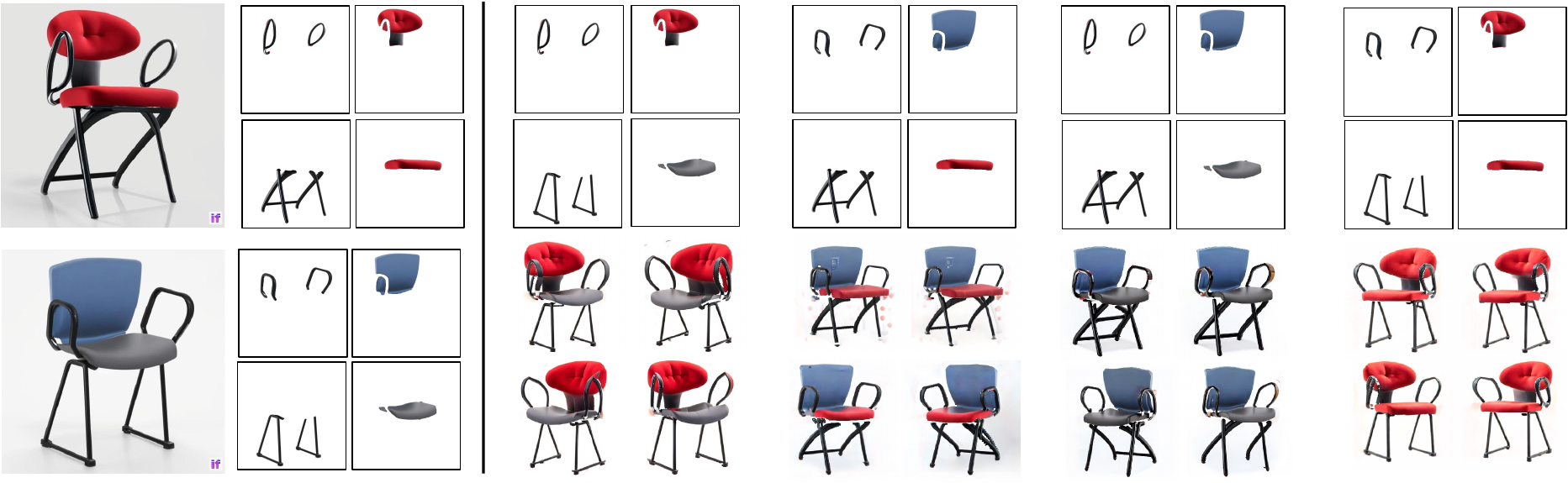}
    \caption{Concept composition results for various compositions of chair parts.}
    \label{fig:results-intra-chairs-2}
\end{figure*}

\paragraph{Chair.} Parts are: \texttt{armrest}, \texttt{backrest}, \texttt{legs}, and \texttt{seat}.  
Fig.~\ref{fig:results-intra-chairs-1} and Fig.~\ref{fig:results-intra-chairs-2} show concept composition results across 6 different chair pairs.

\begin{figure*}[htbp]
    \centering
    \includegraphics[width=\linewidth]{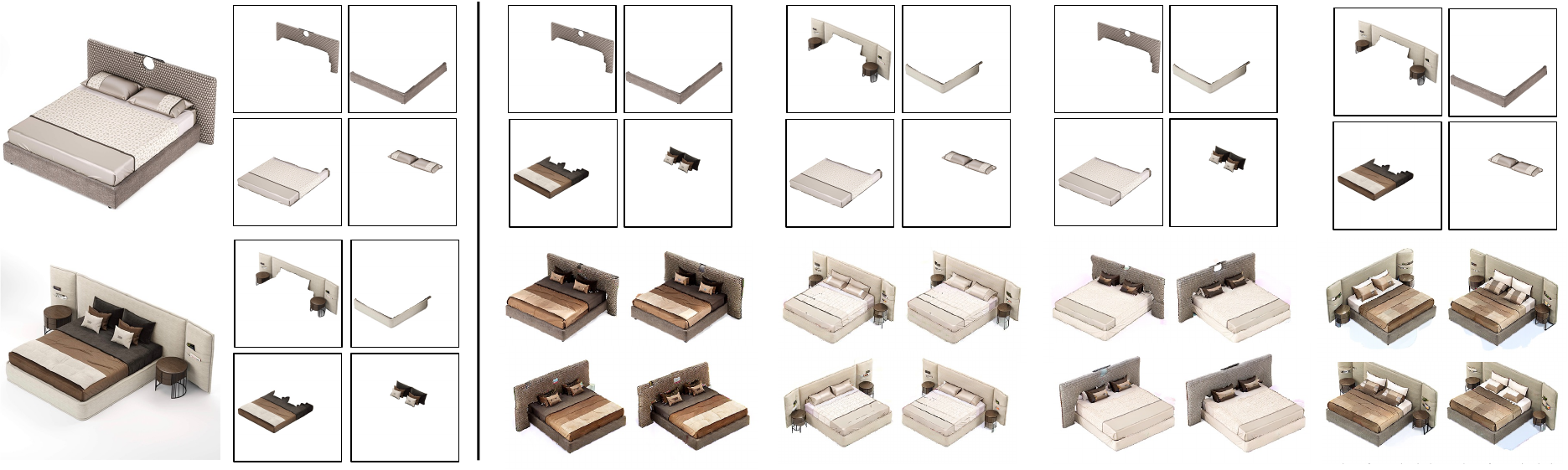}
    \caption{Concept composition results for beds.}
    \label{fig:results-intra-bed}
\end{figure*}

\paragraph{Bed.} Parts are: \texttt{headboard}, \texttt{base}, \texttt{mattress}, and \texttt{pillow}.  
Fig.~\ref{fig:results-intra-bed} shows concept composition for a pair of beds.

\begin{figure*}[htbp]
    \centering
    \includegraphics[width=\linewidth]{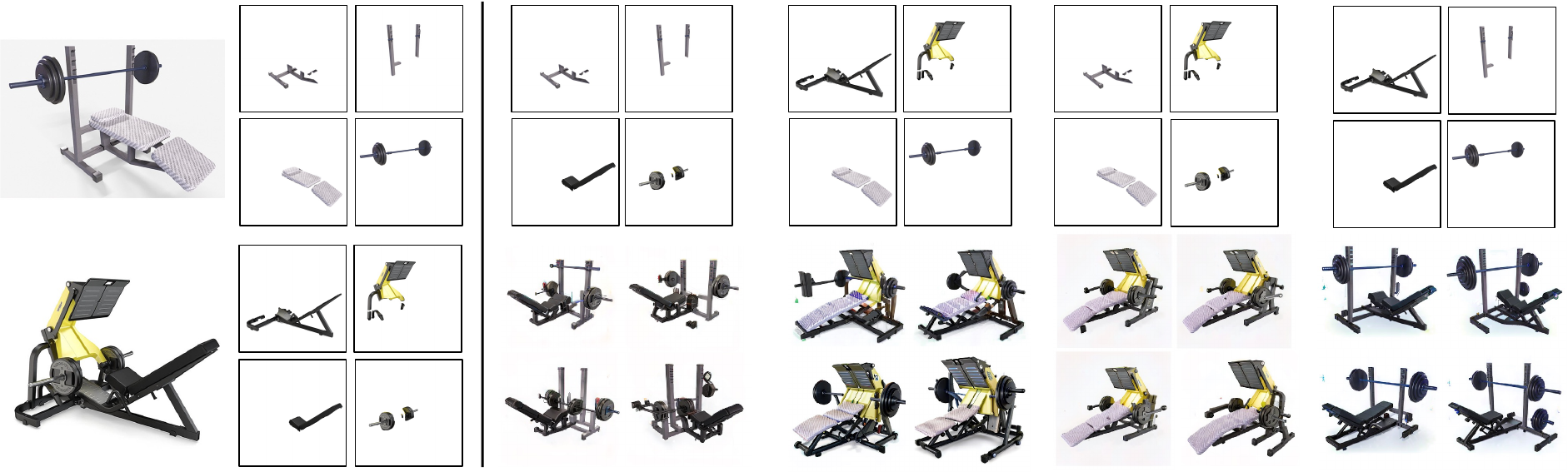}
    \caption{Concept composition results for gym equipments.}
    \label{fig:results-intra-gym}
\end{figure*}

\paragraph{Gym Equipment.} Parts are: \texttt{base}, \texttt{stand}, \texttt{seat}, and \texttt{weight}.  
Fig.~\ref{fig:results-intra-gym} shows concept composition for a pair of gym equipment examples.

\begin{figure*}[htbp]
    \centering
    \includegraphics[width=\linewidth]{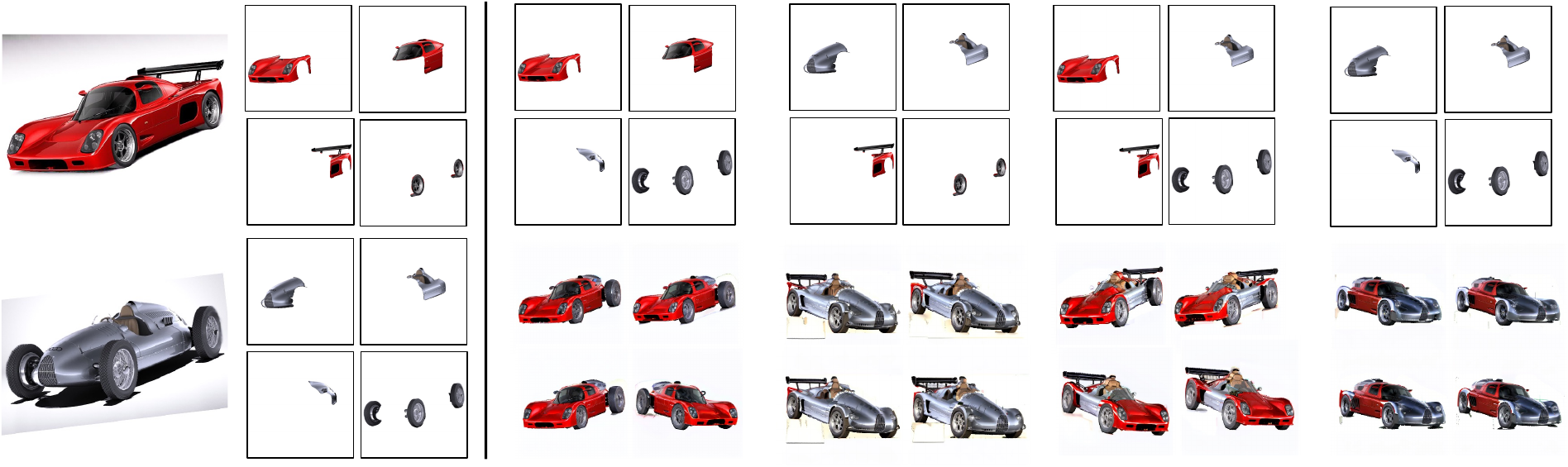}\\
    \hdashrule[0.5ex]{\linewidth}{1pt}{3pt 2pt} \\  
    \includegraphics[width=\linewidth]{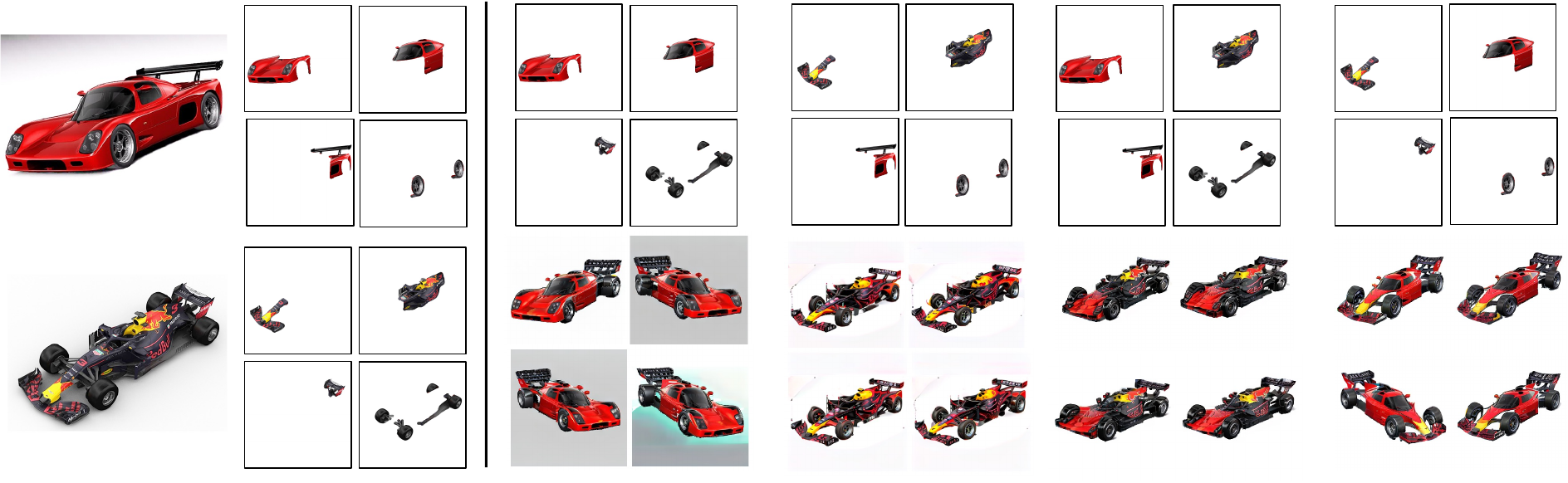}\\
     \hdashrule[0.5ex]{\linewidth}{1pt}{3pt 2pt} \\
     \includegraphics[width=\linewidth]{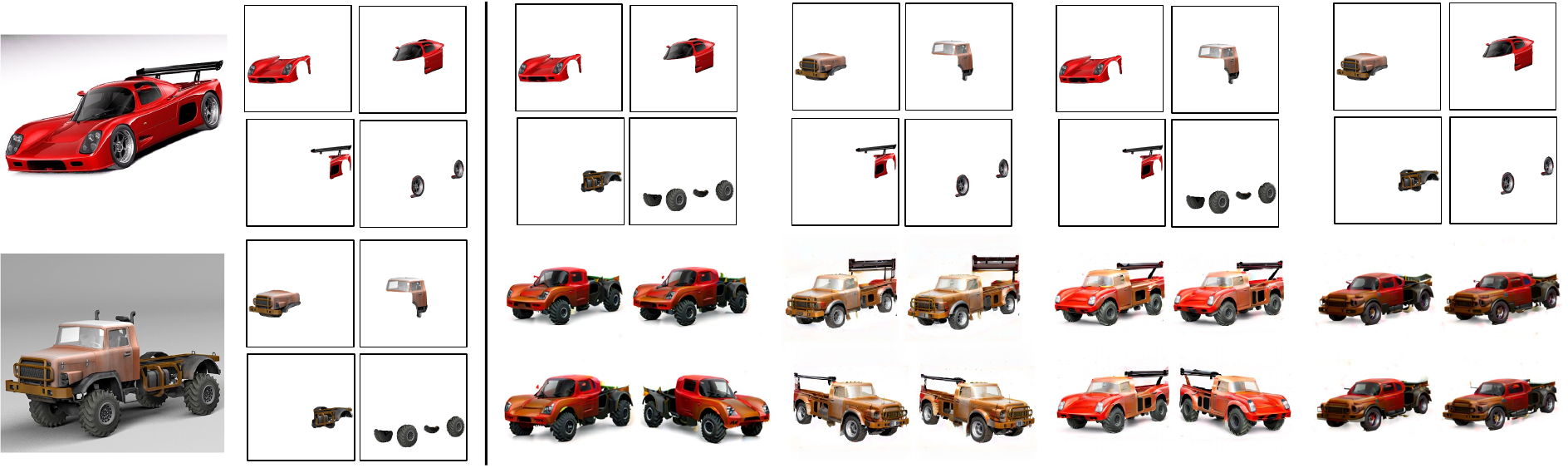}
    \caption{Concept composition results for vehicles. The first pair contains a sports car and an old formula one car. The second pair contains a sports car and a modern formula one car. The last pair contains a sports car and a truck.}
    \label{fig:results-intra-car}
\end{figure*}

\paragraph{Vehicle.} Parts are: \texttt{front}, \texttt{cockpit}, \texttt{tail}, and \texttt{wheels}.  
Fig.~\ref{fig:results-intra-car} shows concept composition for three pairs of vehicles.

\begin{figure*}[htbp]
    \centering
    \includegraphics[width=\linewidth]{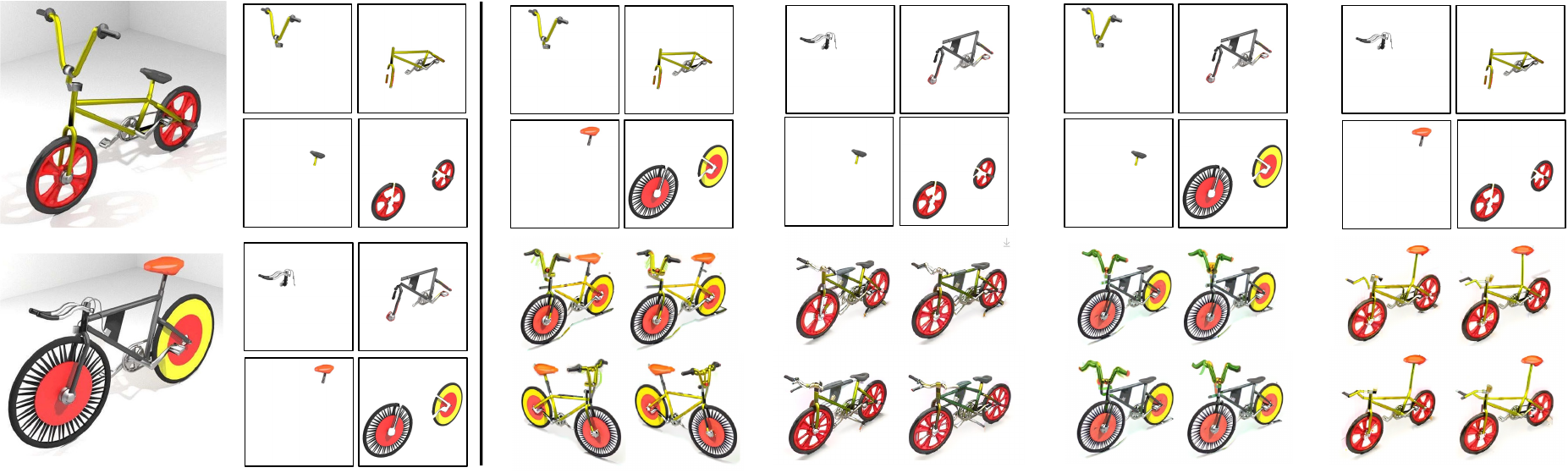}
    \caption{Concept composition results for bikes.}
    \label{fig:results-intra-bike}
\end{figure*}

\paragraph{Bike.} Parts are: \texttt{handle}, \texttt{frame}, \texttt{seat}, and \texttt{wheels}.
Fig.~\ref{fig:results-intra-bike} shows concept composition results for a pair of bikes.

\begin{figure*}[htbp]
    \centering
    \includegraphics[width=\linewidth]{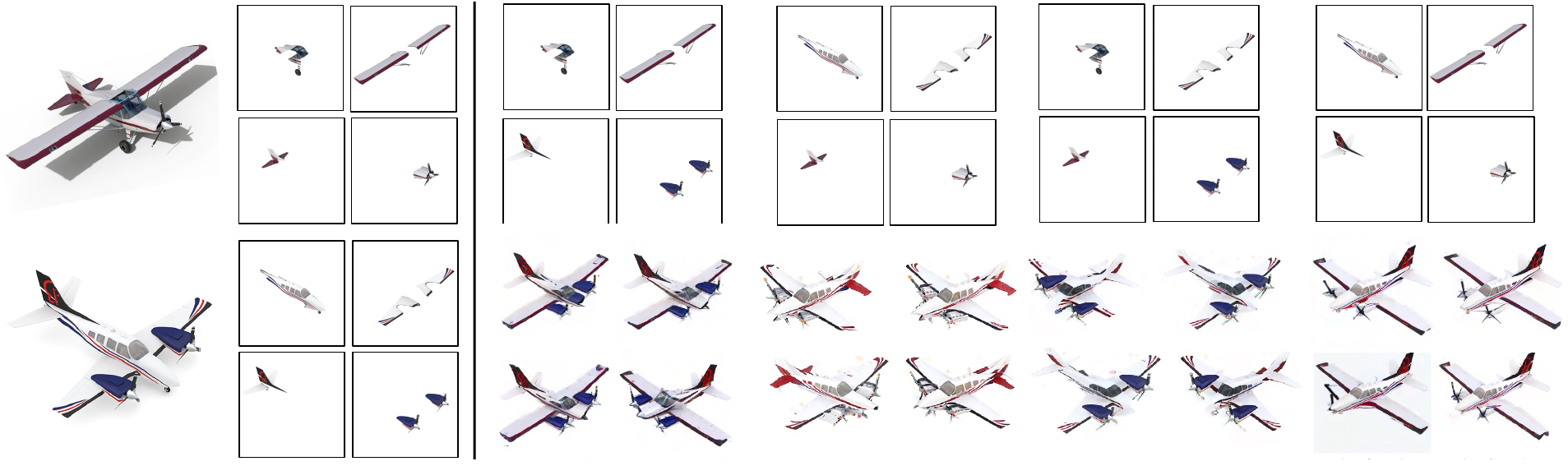}
    \caption{Concept composition results for planes.}
    \label{fig:results-intra-plane}
\end{figure*}

\paragraph{Plane.} Parts are: \texttt{body}, \texttt{wing}, \texttt{tail}, and \texttt{engine}.  
Fig.~\ref{fig:results-intra-plane} shows concept composition results for a pair of planes.

\begin{figure*}[htbp]
    \centering
    \includegraphics[width=\linewidth]{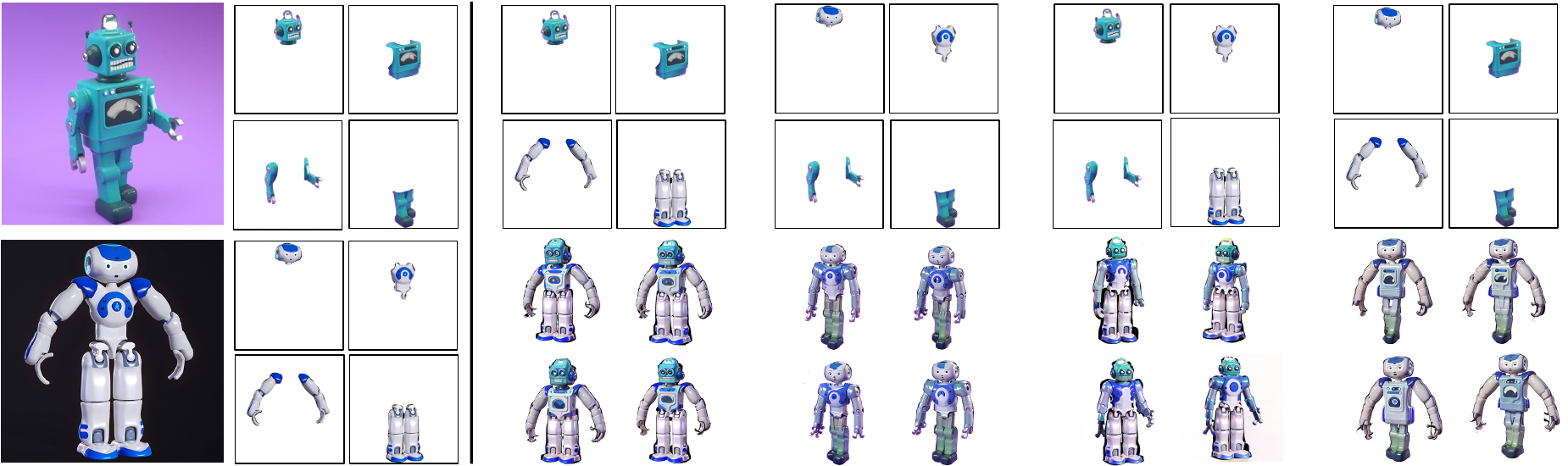}
    \caption{Concept composition results for robots.}
    \label{fig:results-intra-robot}
\end{figure*}

\paragraph{Robot.} Parts are: \texttt{head}, \texttt{body}, \texttt{arm}, and \texttt{leg}.  
Fig.~\ref{fig:results-intra-robot} shows concept composition results for a pair of robots.

\begin{figure*}[htbp]
    \centering
    \includegraphics[width=\linewidth]{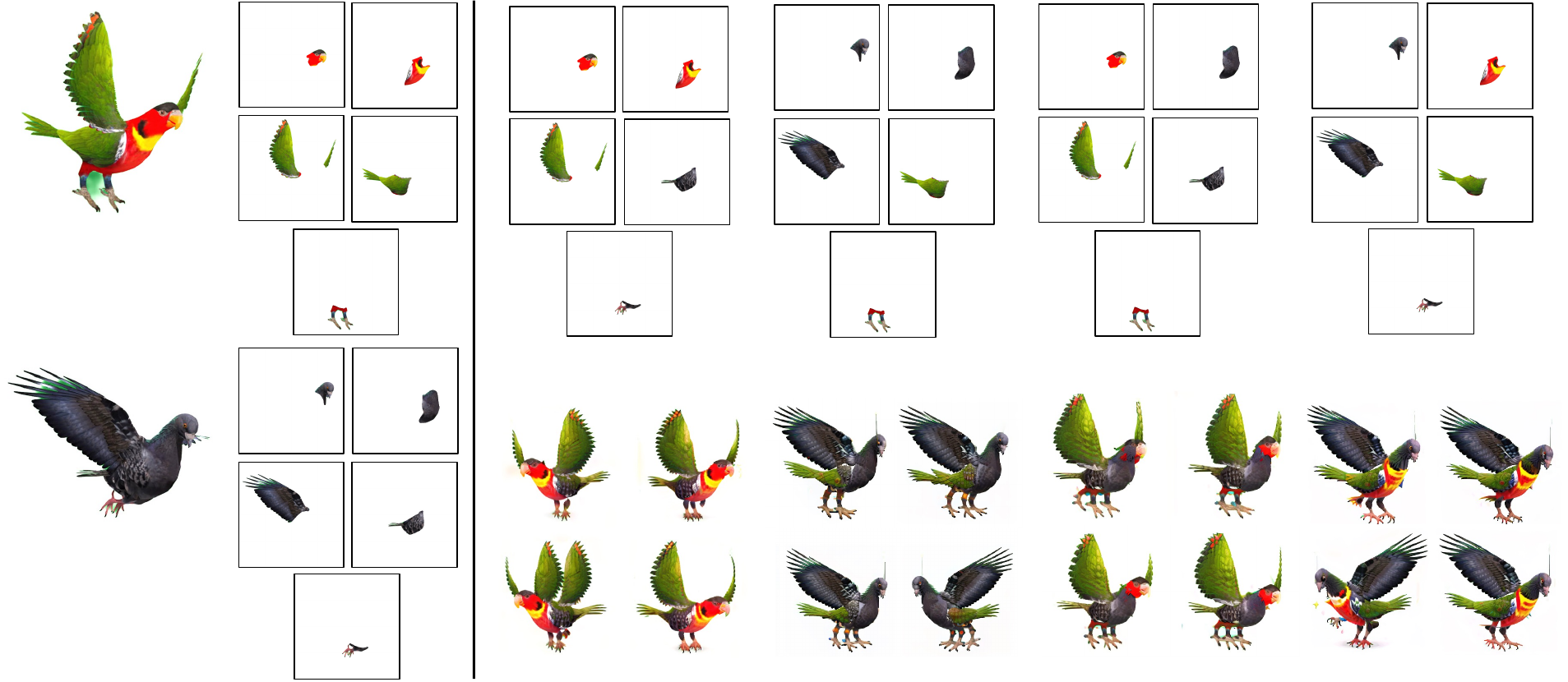}\\
    \hdashrule[0.5ex]{\linewidth}{1pt}{3pt 2pt} \\  
    \includegraphics[width=\linewidth]{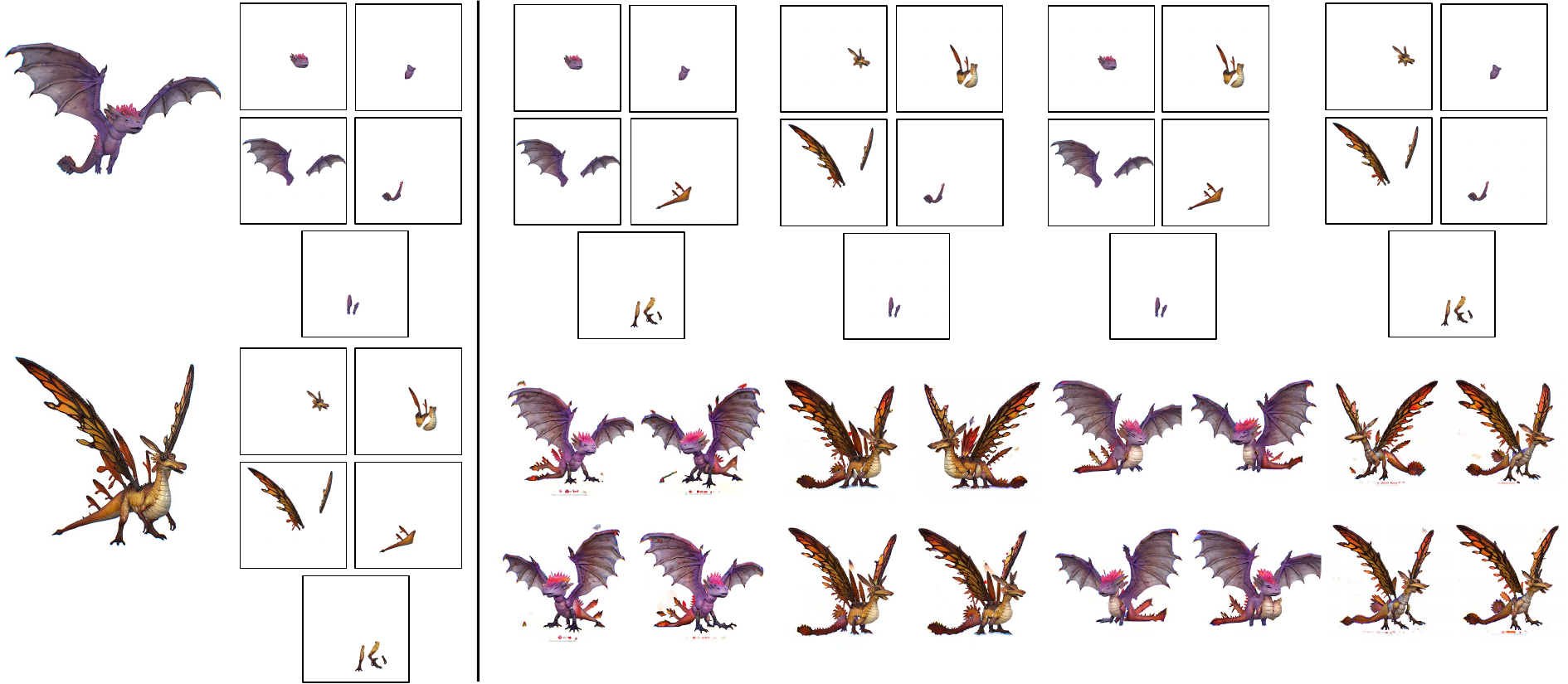}\\
    \caption{Concept composition results for birds and dragons.}
    \label{fig:results-intra-bird-creature}
\end{figure*}

\paragraph{Bird and Dragon.} Parts are: \texttt{head}, \texttt{body}, \texttt{wings}, \texttt{tail}, and \texttt{legs}.  
Fig.~\ref{fig:results-intra-bird-creature} shows concept composition for a bird and a creature pair.

\begin{figure*}[htbp]
    \centering
    \includegraphics[width=0.9\linewidth]{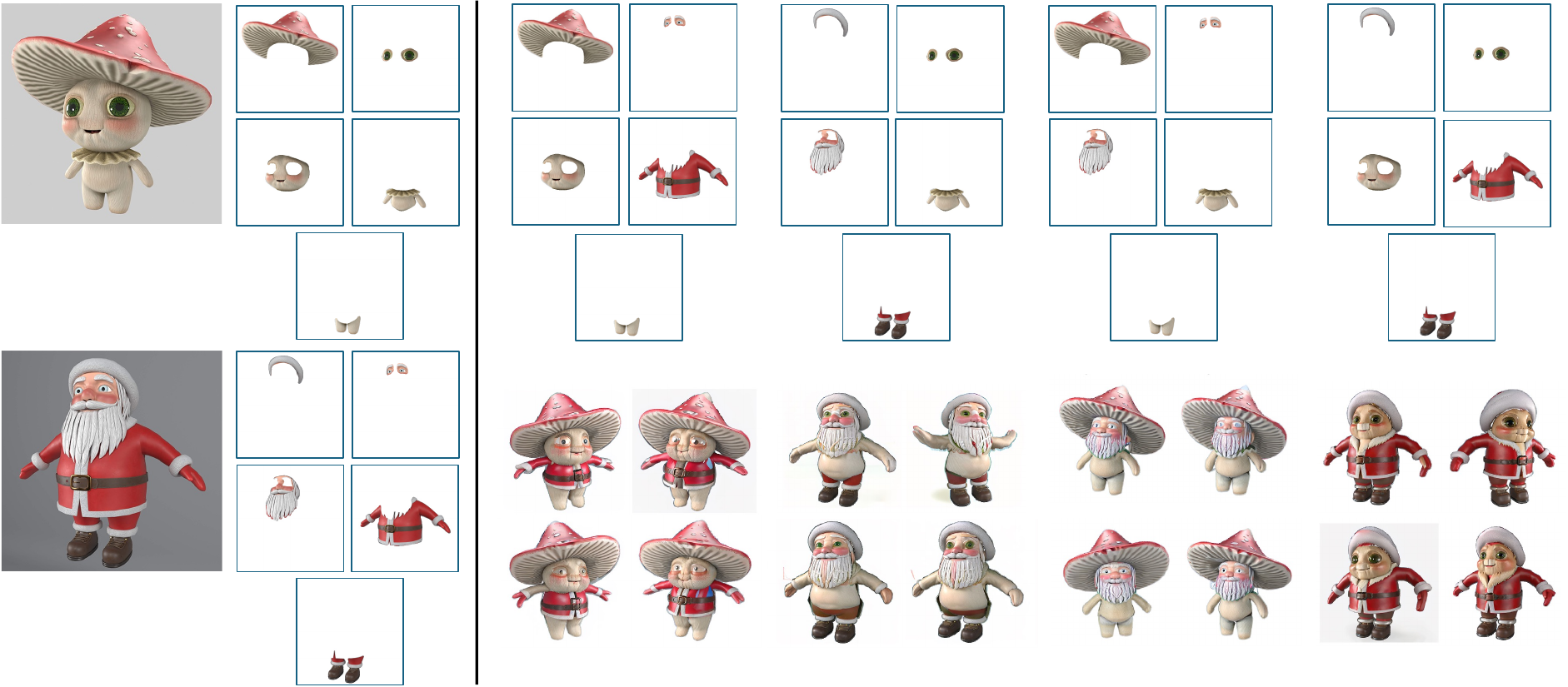} \\
    \hdashrule[0.5ex]{0.9\linewidth}{1pt}{3pt 2pt} \\  
    \includegraphics[width=0.9\linewidth]{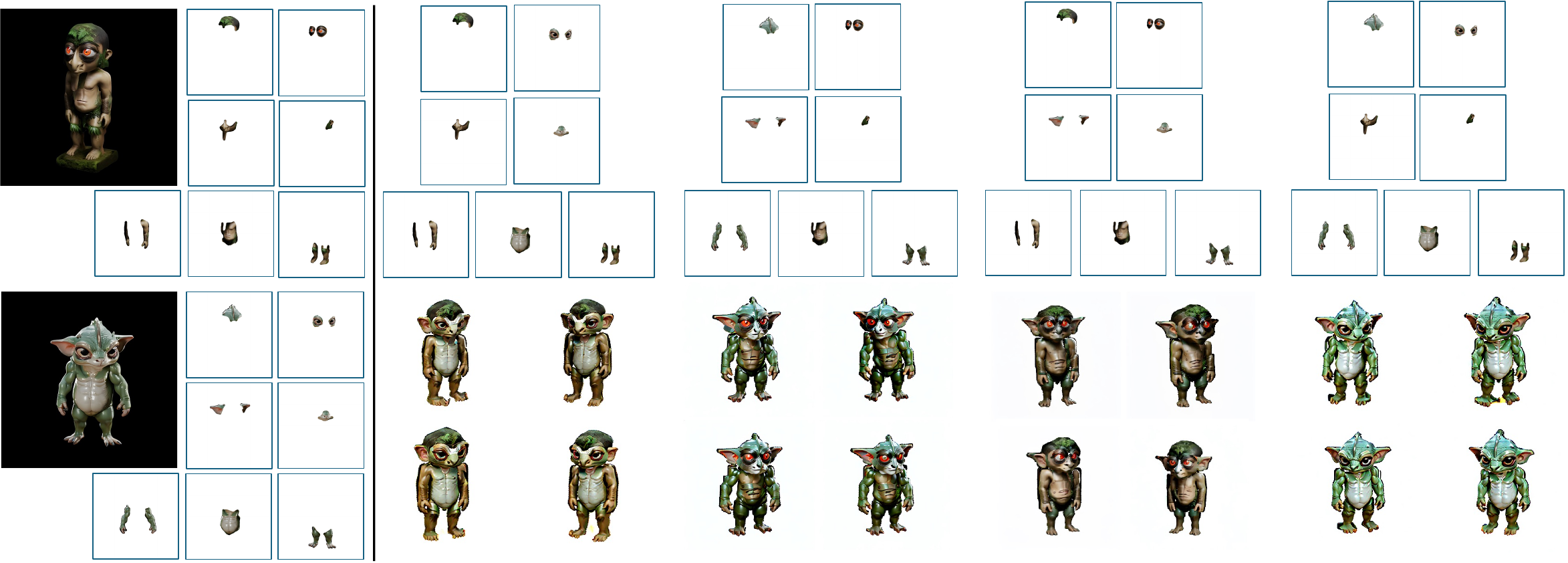} \\
    \hdashrule[0.5ex]{0.9\linewidth}{1pt}{3pt 2pt} \\  
    \includegraphics[width=0.9\linewidth]{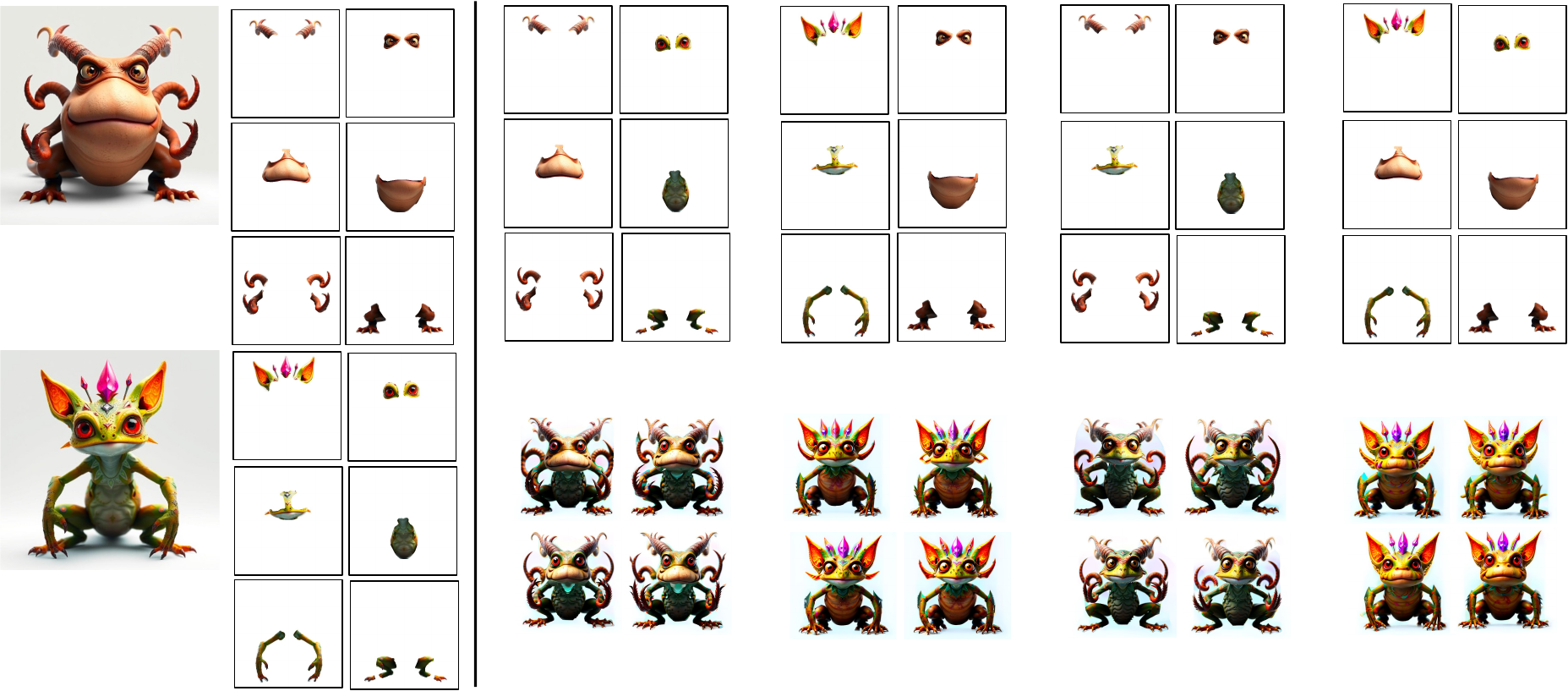} \\
    \caption{Concept composition results for virtual creatures: a mushroom-like character and a santa, a hermit and a reptile, and two virtual creatures.}
    \label{fig:results-intra-character-1}
\end{figure*}

\paragraph{Virtual Characters.} Fig.~\ref{fig:results-intra-character-1} shows concept composition for a mushroom-like character and a santa-like character. Parts are: \texttt{head}, \texttt{eyes}, \texttt{face}, \texttt{body}, and \texttt{legs}.
Fig.~\ref{fig:results-intra-character-1} shows concept composition for a hermit-like character and a reptile-like character. This example aim to demonstrate that our method can learn a large number (i.e., in this case, 7 parts per image) of fine-grained concepts for parts and re-compose them. Parts are: \texttt{head}, \texttt{eyes}, \texttt{face}, \texttt{nose}, \texttt{arms}, \texttt{body}, and \texttt{legs}. 

\subsection{Cross-category Results}
\label{append:subsec:cross-category}

\begin{figure*}[htbp]
    \centering
    \includegraphics[width=0.9\linewidth]{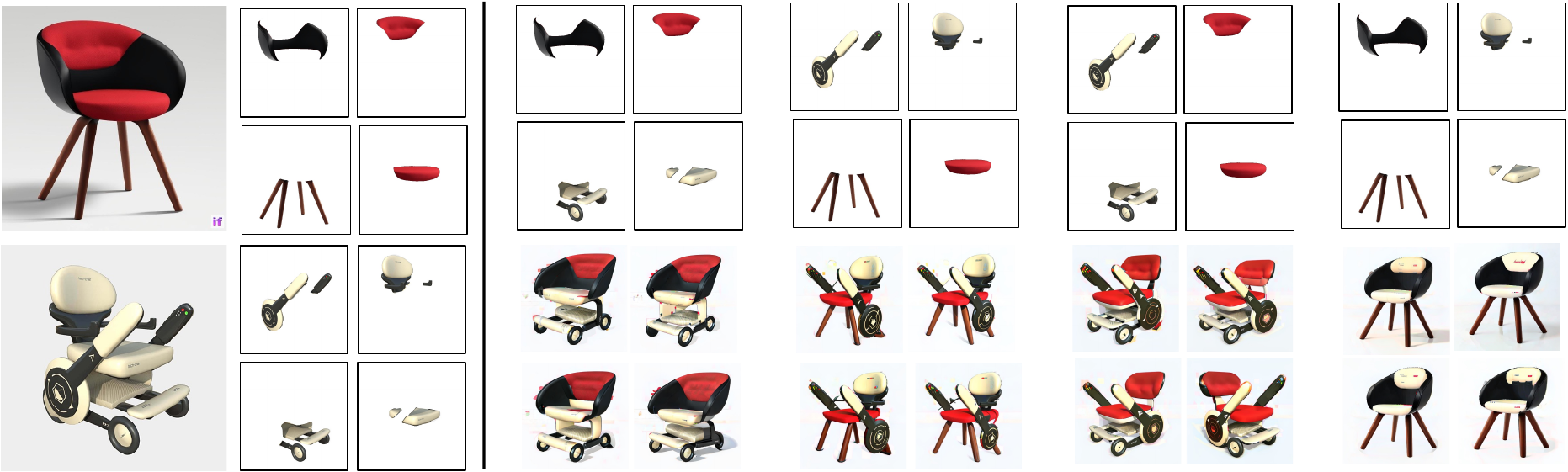}\\
     \hdashrule[0.5ex]{0.9\linewidth}{1pt}{3pt 2pt} \\  
    \includegraphics[width=0.9\linewidth]{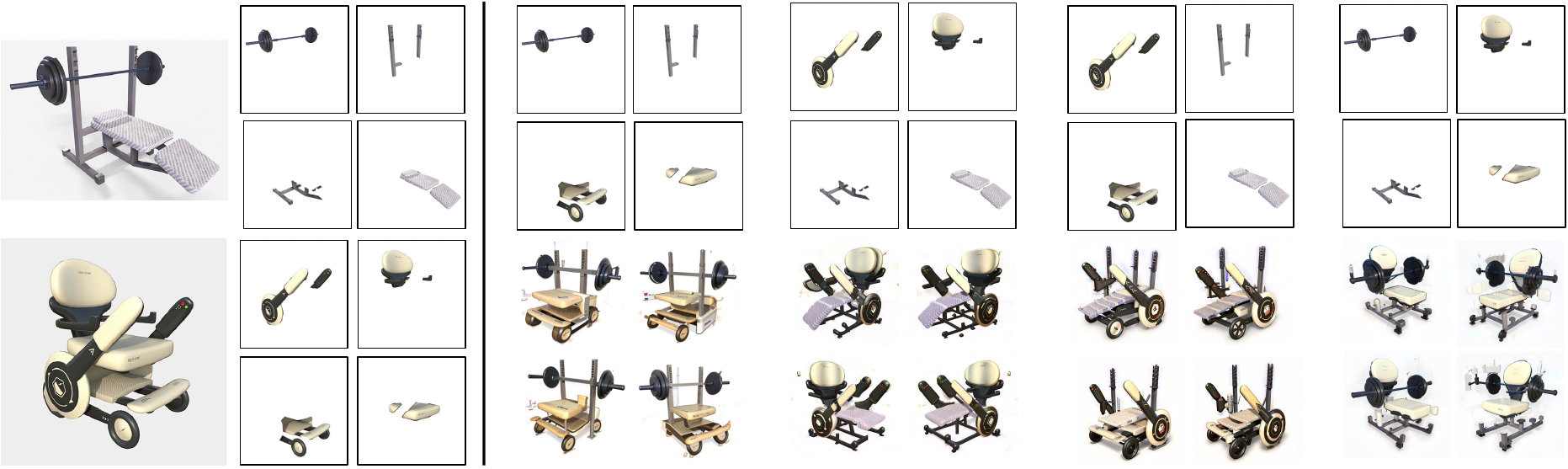}\\
     \hdashrule[0.5ex]{0.9\linewidth}{1pt}{3pt 2pt} \\  
    \includegraphics[width=0.9\linewidth]{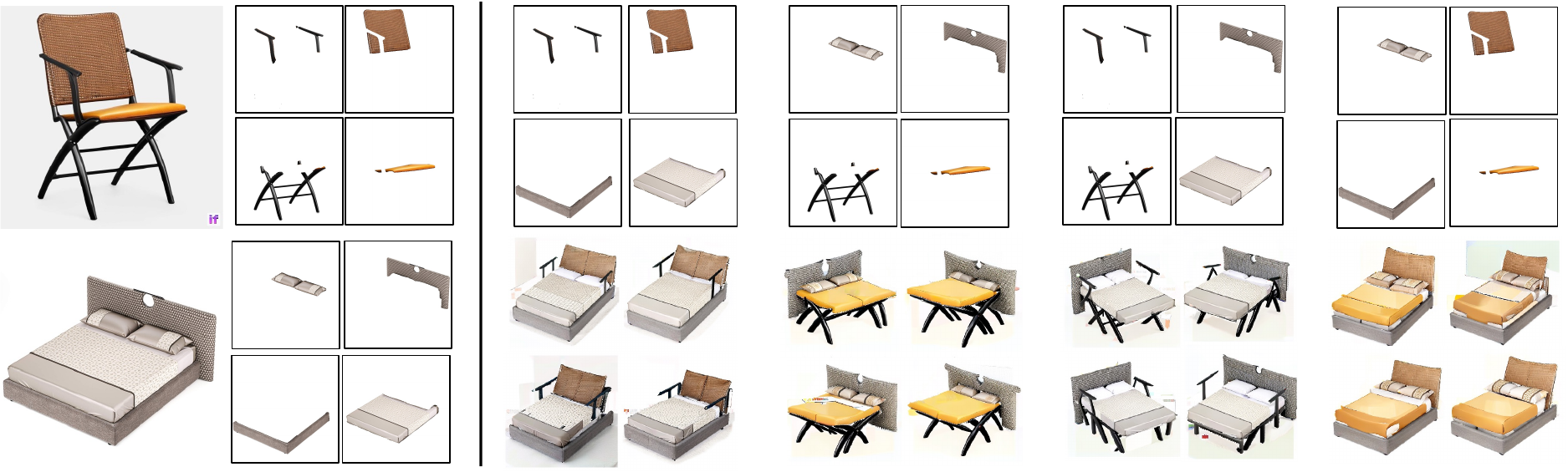}\\
     \hdashrule[0.5ex]{0.9\linewidth}{1pt}{3pt 2pt} \\  
     \includegraphics[width=0.9\linewidth]{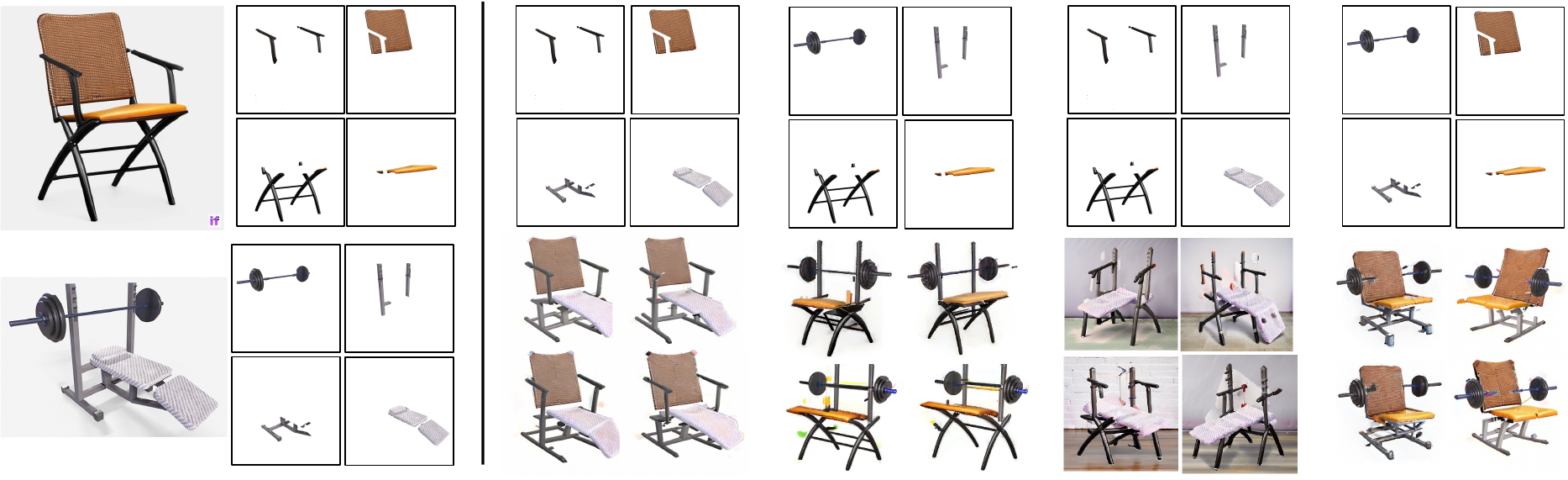}\\
    \caption{Concept composition results for cross-category examples: a chair and a wheel chair, a gym equipement and a wheel chair, a chair and a bed, and a chair and a gym equipment.}
    \label{fig:results-cross}
\end{figure*}

To demonstrate generalization beyond intra-category recomposition, we evaluate our method on cross-category part composition. Fig.~\ref{fig:results-cross} shows 4 hybrid compositions: a chair and a wheel chair, a gym equipement and a wheel chair, a chair and a bed, and a chair and a gym equipment..

\end{document}